\documentclass[aps,reprint,prb,letterpaper,superscriptaddress]{revtex4-2}
\usepackage{lmodern}

\pdfoutput=1

\usepackage{amsmath,amssymb,amsthm,mathrsfs, cancel,comment}
\usepackage{siunitx,mhchem,nicefrac}

\usepackage{miller,physics,commath,hyperref}

\usepackage{graphicx,wrapfig,subfig,ragged2e}

\usepackage{tabularx, booktabs,blindtext,array,dcolumn}

\usepackage{hyperref}
\hypersetup{
	colorlinks=true,
	linkcolor=blue,
	citecolor=blue,
	urlcolor=blue,
}

\usepackage{lipsum}
\usepackage{xcolor}
\usepackage[figuresright]{rotating}

\usepackage[all]{nowidow}
\usepackage[utf8]{inputenc}
\usepackage{bbm,geometry}
\usepackage{float,blkarray}

\graphicspath{{images/}}

\begin{document}

\title{Anisotropy reduction and tunability of hole-spin qubit g-factor in strained parabolic Ge/SiGe quantum wells}

\author{R.\ K.\ L.\ Colmenar}
\email{ralphkc1@umd.edu}
\email{rkcolme@lps.umd.edu}
\affiliation{Department of Physics,
	University of Maryland,
	College Park, Maryland 20742, USA
}
\affiliation{
	Laboratory for Physical Sciences,
	College Park, Maryland 20742, USA
}

\author{Arthur Lin}
\email{artlin@umd.edu}
\email{arthur.lin@nist.gov}
\affiliation{Department of Physics,
	University of Maryland,
	College Park, Maryland 20742, USA
}
\affiliation{Nanoscale Device Characterization Division,
	National Institute of Standards and Technology,
	Gaithersburg, Maryland 20899, USA
}

\author{Omadillo Abdurazakov}
\affiliation{Department of Physics,
	University of Texas at El Paso,
	El Paso, Texas 79968, USA
}

\author{Yun-Pil Shim}
\affiliation{Department of Physics,
	University of Texas at El Paso,
	El Paso, Texas 79968, USA
}

\author{Garnett W.\ Bryant}
\affiliation{Nanoscale Device Characterization Division,
	National Institute of Standards and Technology,
	Gaithersburg, Maryland 20899, USA
}

\author{Charles Tahan}
\affiliation{Department of Physics,
	University of Maryland,
	College Park, Maryland 20742, USA
}


\begin{abstract}
	Hole-spin qubits in planar Ge/SiGe heterostructures have attracted significant attention in recent years owing to their favorable electrical characteristics and prolonged coherence times. However, the strong spin-orbit interaction also makes them susceptible to charge noise and inhomogeneous strain. This is further exacerbated by the highly anisotropic g-factor of the planar design. Although there are some known strategies to suppress charge noise, one approach is to engineer an isotropic g-factor. In this work we analyze how qubit confinement profile affects the g-factor of hole-spin qubits. We show that decreasing the characteristic in-plane qubit confinement length reduces the g-factor anisotropy. We perform analytical and numerical analysis to compare two types of quantum wells: square wells and parabolic wells. We show that square wells have limited tunability, while parabolic wells offer broader tunability, making them more promising for qubit engineering.
\end{abstract}

\maketitle

%%%%%%%%%%%%%%%%%%%%%%%%%%%%%%%%%%%%%%%%%%%%%%%%%%%
\begin{figure}[hb]
    \centering
    \subfloat[20\,nm lateral confinement]{
    \includegraphics[width=\linewidth]{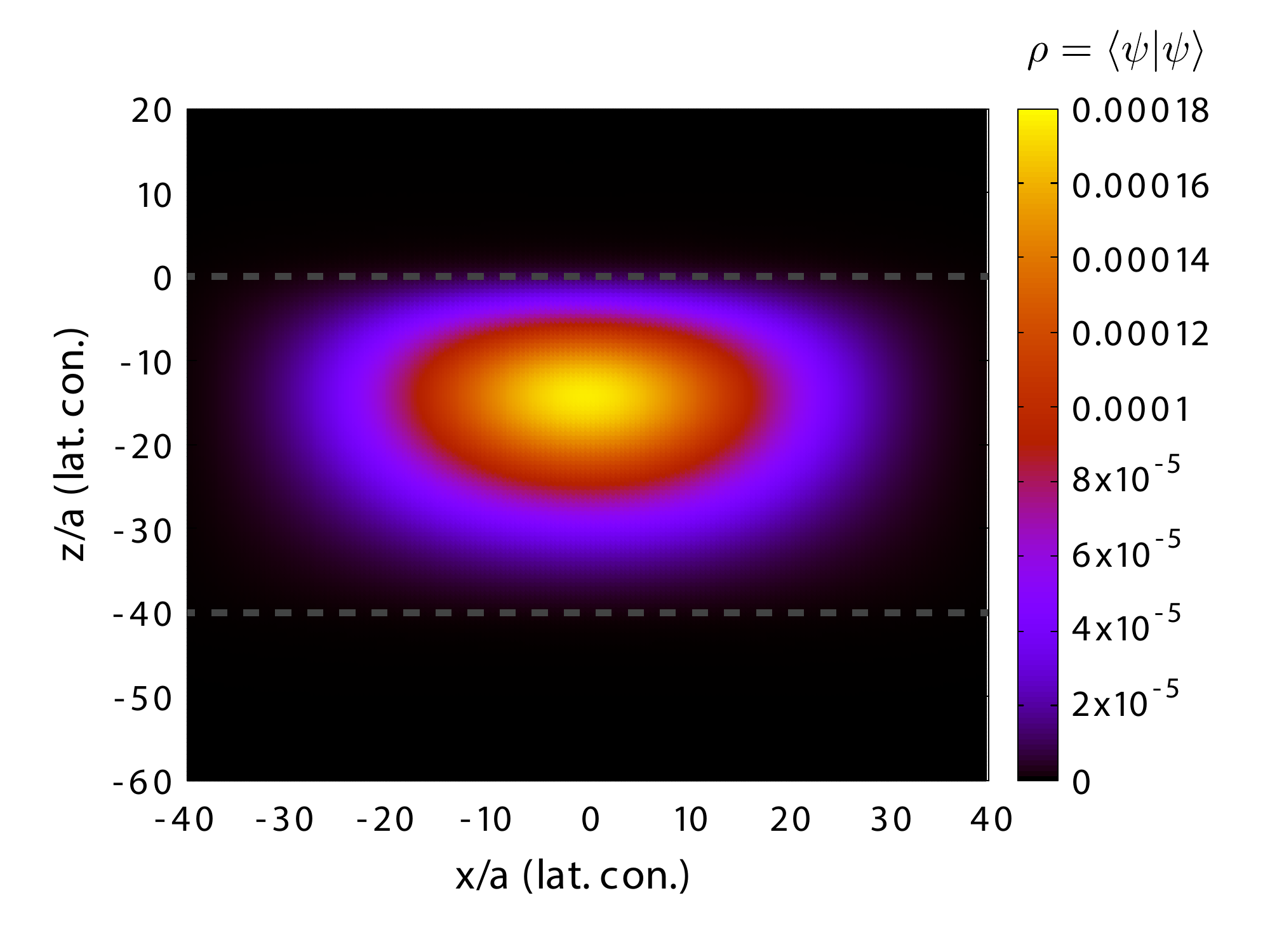}
    }\\
    \subfloat[5\,nm lateral confinement]{
    \includegraphics[width=\linewidth]{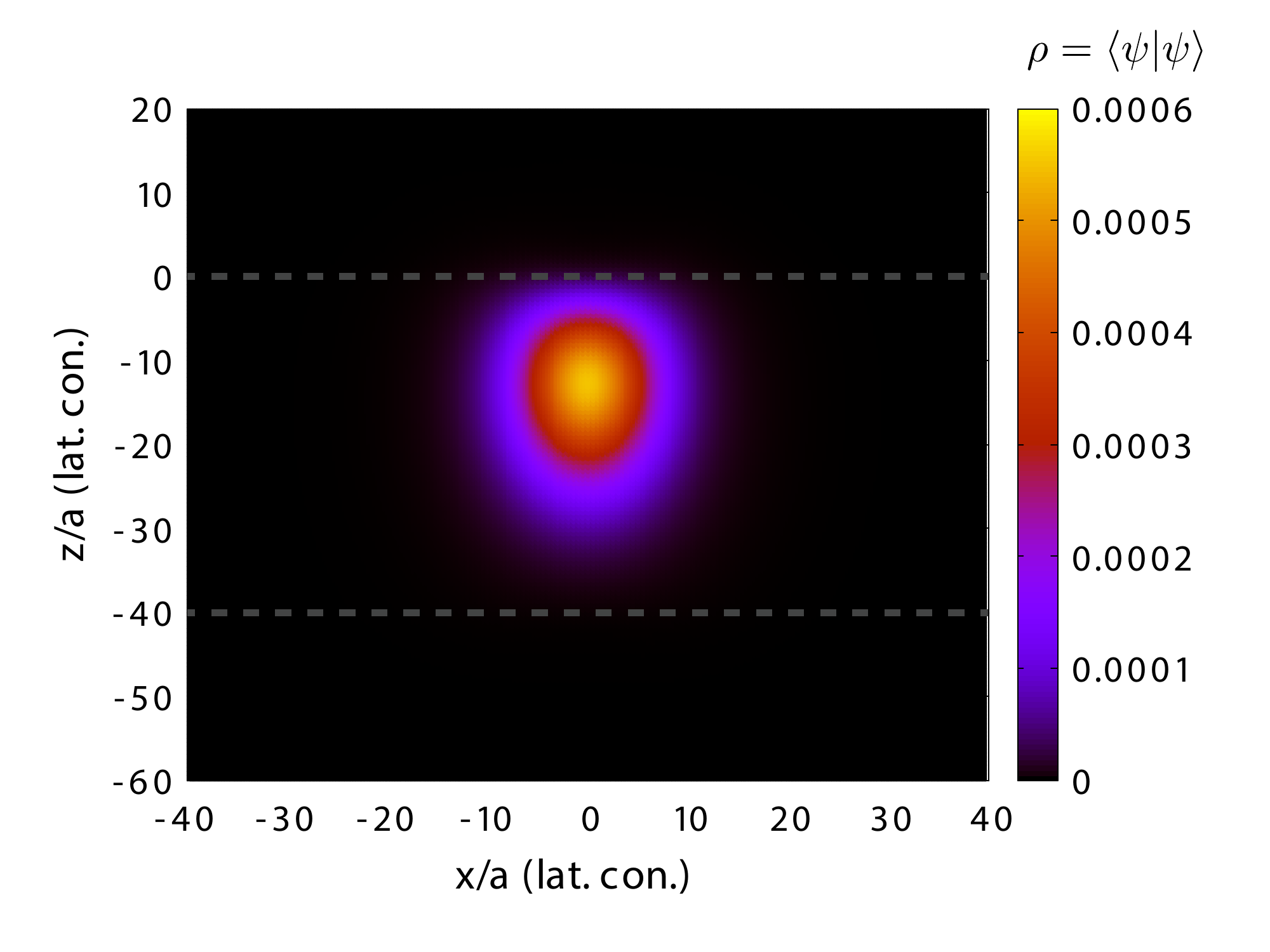}
    }
    \caption{\justifying \small
    A plot of the ground state hole wavefunction probability (charge density) along the x-z~plane for a $40$ lattice constant ($a=a_{\ce{Si_{0.2}Ge_{0.8}}}=0.5613~\text{nm}$) thick quantum well with (a) a $20\,\text{nm}$ and (b) a $5\,\text{nm}$ full-width-half-maximum lateral confinement. The anisotropic nature of the hole effective mass skews the wavefunction along one direction. This asymmetry can be made symmetric by tuning the confinement strength which, intuitively, should mitigate the anisotropy in certain geometry-sensitive properties such as the effective hole g-factor.}
    \label{fig:wf}
\end{figure}

\section{Introduction}
Semiconductor quantum dots are among the leading platforms for realizing qubits due to their long coherence times and compatibility with established semiconductor fabrication processes. Whereas the usual paradigms have focused on trapping electron spins, hole spins in strained Ge/SiGe heterostructures~\cite{Watzinger_2018,Hendrickx_2020_1,Hendrickx_2020_2,Bosco_2021,Bosco_2021_2,Froning_2021,Hendrickx_2021,Jirovec_2021,Mutter_2021,Scappucci_2021,Studenikin_2021,Terrazos_2021,Wang_2021,Adelsberger_2022,Assali_2022,Bosco_2022,Jirovec_2022,Wang_2022,Abadillo-Uriel_2023,Fang_2023,Sarkar_2023,Wang_2023,Borsoi_2024,Hendrickx_2024,Mauro_2025,Wang_2025} have emerged as attractive alternatives owing to their low in-plane effective mass, lack of valley degeneracy at the $\Gamma$ point, and strong electrically tunable spin-orbit interaction (SOI)~\cite{Scappucci_2021,Studenikin_2021,Fang_2023}. External electric fields can induce an effective Rashba SOI, enabling in situ tunability of qubit properties and all-electrical control of hole spins via electric-dipole spin resonance. Moreover, the $p$-symmetry of valence holes suppresses the contact hyperfine interaction with the host semiconductor's nuclear spins~\cite{Kolodrubetz_2009,Brunner_2009,Fallahi_2010} which, in addition to isotopic purification of Ge~\cite{Itoh_1993}, improves hole-spin coherence times~\cite{Tyryshkin_2012}. Considerable experimental efforts have demonstrated high-fidelity logical gates in both single-spin~\cite{Hendrickx_2020_1, Hendrickx_2020_2} and singlet-triplet~\cite{Jirovec_2021,Jirovec_2022,Wang_2023} encodings, which have been scaled up to multi-qubit arrays~\cite{Hendrickx_2021,Borsoi_2024} within just a few years.

Conversely, the strong SOI can also be detrimental to the qubit quality since it couples the hole-spin to electrostatic fluctuations~\cite{Wang_2025} and strain inhomogeneities~\cite{Abadillo-Uriel_2023}. Local charge traps, alloy disorder, and growth- or cooldown-induced strain impart random shifts to the qubit quantization axis which can lead to dephasing. Small variations in the electrostatic confinement are amplified by the highly anisotropic g-factor which can measurably shift Rabi frequencies~\cite{Abadillo-Uriel_2023} and preclude charge-noise sweet spots~\cite{Hendrickx_2024}. The large anisotropy is caused by the highly asymmetric dot confinement, where lateral dimensions are typically $\geq 100\,\text{nm}$, while the quantum well thickness is often $<30\,\text{nm}$.

Thus, one might address the aforementioned issues by reducing the underlying anisotropy of the hole-spin qubit, i.e., by engineering the qubit to be more isotropic (see Fig.~\ref{fig:wf}). A nearly isotropic ground state can be achieved by using a thicker quantum well and a much stronger lateral confinement. A key challenge, however, is the structural limit imposed by strain. In conventional square well heterostructures, the sharp strain gradient sets an upper limit to the allowable material thickness while maintaining lattice stability~\cite{Schaffler_1997}. This motivates the use of alternative quantum well architectures such as parabolic SiGe wells with smoothly graded Ge concentration~\cite{Ballabio_2019}. Unlike square wells, parabolic wells distribute strain more uniformly and enable significantly thicker wells, leading to more extended vertical confinement.

In this work, we explore the effective hole-spin qubit g-factor and its anisotropy in planar Ge/SiGe quantum dots, with emphasis on how confinement geometry influences g-factor behavior. We begin in Sec.~\ref{sec:model} by constructing detailed models of both square and parabolic quantum wells, including strain and electrostatic confinement. Sec.~\ref{sec:well-comparison} analyzes how well geometry affects vertical confinement of the hole wavefunction and shows analytically that, in the presence of a constant electric field along the growth direction of the heterostructure, square wells impose a limit on qubit width while parabolic wells do not. We show the difference between the g-factor properties of the two well types and provide predictions on the necessary confinement lengths to reduce the g-factor anisotropy. Finally, we present a detailed analysis of our numerical g-factor simulations in Sec.~\ref{subsec:results}. Our results show that parabolic wells offer a more tunable g-factor than square wells, making them a promising platform for qubit engineering.

%%%%%%%%%%%%%%%%%%%%%%%%%%%%%%%%%%%%%%%%%%%%%%%%%%%

\section{Modeling the well}
\label{sec:model}
\begin{figure}
    \centering
    \includegraphics[width=\linewidth]{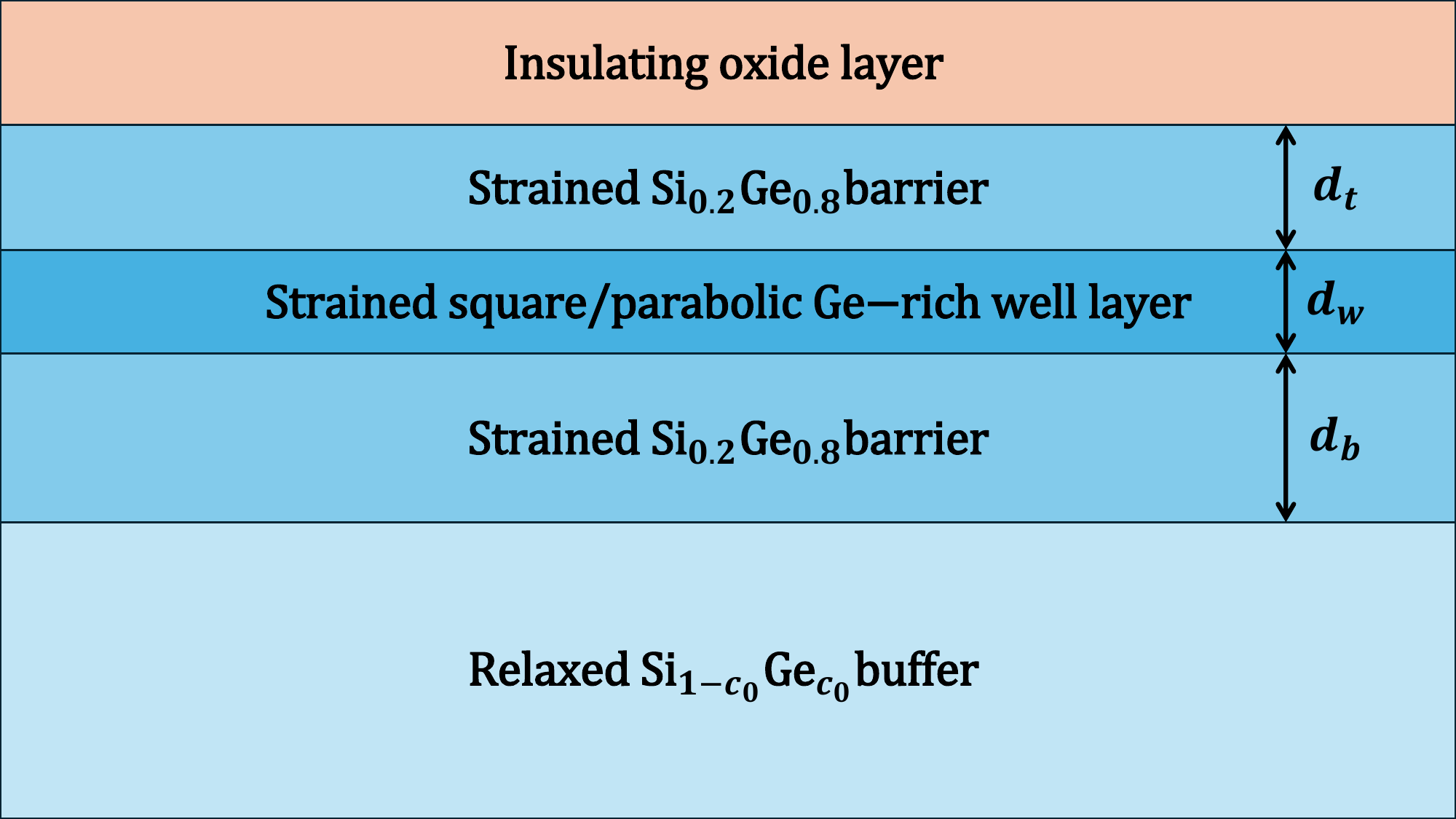}
    \caption{\justifying \small
    A schematic diagram of the heterostructure. A Ge-rich layer, which can be either a square or a parabolic well, is embedded between $\ce{Si_{0.2}Ge_{0.8}}$ barriers and grown atop a relaxed $\ce{Si_{1-c_0}Ge_{c_0}}$ buffer layer. The structure is capped by an insulating oxide layer, above which voltage gates are placed to provide additional electrostatic confinement to produce the dot.}
    \label{fig:schematic}
\end{figure}

We consider a single-hole spin qubit confined within a planar Ge/SiGe heterostructure, where the vertical confinement is provided by either a square or a parabolic quantum well as shown schematically in Fig.~\ref{fig:schematic}. In both configurations, the quantum well consists of a Ge-rich layer embedded between $\ce{Si_{0.2}Ge_{0.8}}$ barriers and grown atop a relaxed (and/or counter-strained) $\ce{Si_{1-c_0}Ge_{c_0}}$ buffer layer. We model the quantum well by expressing physical quantities in terms of the Ge concentration along the growth direction~(z) where the origin is set at the top of the well:
\begin{align}
\label{eq:ge-concentration-square}
    &c_\text{sq}(z) = 
\begin{cases}
        c_0 & -\infty < z < -d_w - d_b,\\
        0.8 & -d_w-d_b < z < -d_w,\\
        1 & -d_w \leq z \leq 0,\\
        0.8 & 0 < z \leq d_t,\\
\end{cases}
\end{align}
and
\begin{align}
\label{eq:ge-concentration-para}
    &c_\text{para}(z) = \nonumber\\
    &
\begin{cases}
        c_0 & -\infty < z < -d_w - d_b,\\
        0.8 & -d_w-d_b < z < -d_w,\\
        -0.2\frac{(z+d_w/2)^2}{d_w^2/4} + 1 & -d_w \leq z \leq 0,\\
        0.8 & 0 < z \leq d_t,\\
\end{cases}
\end{align}
where $c_0$ is the Ge concentration in the buffer layer, $d_{t(b)}$ is the thickness of the top (bottom) barrier, and $d_w$ is the thickness of the well. In this work, we set $d_t = d_b = 50\,\text{nm}$ and leave $d_w$ as a free parameter. An insulating oxide layer caps the structure which we model as an infinite potential barrier for simplicity. Henceforth, physical quantities related to the quantum well are assumed to take concentration values such as Eqs.~\eqref{eq:ge-concentration-square} and \eqref{eq:ge-concentration-para} for their input. For example, we denote the composition-dependent valence band edge of the heterostructure as $\text{VBE}[c_\text{sq/para}(z)]$, or $\text{VBE}[c]$, for a fixed concentration of $\ce{Si_{1-c}Ge_{c}}$.

Strain is another important aspect that needs to be included in the model since it affects the valence band edge and the splitting between heavy and light hole states, thereby influencing the hole's confinement and effective spin properties. We ensure the mechanical stability of the entire heterostructure by imposing a strain-balancing constraint on the buffer layer concentration $c_0$. The strain in the heterostructure is obtained through the relaxed lattice constant which we model as
\begin{equation}
\label{eq:lattice-constant}
    a[c] = a_\text{Si}(1-c)+ a_\text{Ge}c - b c (1-c),
\end{equation}
where $a_\text{Si}=\SI{0.5431}{\nano\meter}$ and $a_\text{Ge}=\SI{0.5658}{\nano\meter}$ are the relaxed Si and Ge lattice constants, respectively, and $b = \SI{0.0027}{\nano\meter}$ is a bowing parameter. The strain in the growth plane of the heterostructure relative to the relaxed buffer is 
\begin{equation}
\label{eq:strain}
    \varepsilon[c] = -1 + a[c]/a[c_0],
\end{equation}
which is negative (positive) when the material is compressive (tensile) strained. Assuming that the elastic stiffness constant of SiGe materials is roughly constant, the strain-balancing condition can be reduced to the average lattice method~\cite{Harrison_2016}:
\begin{align}
\label{eq:strain-balancing}
    a[c_0] = \frac{a[c_b](d_b+d_t) + \int_\text{well}a[c(z)]\mathrm{d}z}{d_b+d_t+d_w},
\end{align}
where the integral is taken over the well region. This constraint defines the value of $c_0$ that prevents the formation of dislocations by matching the average lattice constant of the heterostructure to that of the buffer.  In this work, we consider two strain cases: the first case ignores strain balancing which means the buffer is identical to the barrier (i.e., $c_0 = 0.8$), while the second case is when we include strain balancing which requires solving Eq.~\eqref{eq:strain-balancing} for $c_0$. 

Furthermore, voltage gates are placed above the oxide layer and serve to provide additional electrostatic confinement for the dot. To simplify our considerations, we assume that gate-defined potentials are separable into an in-plane and an out-of-plane component.  In the out-of-plane direction, the voltage gate applies a constant electric field bias in the well: $V_\text{bias}(z) = -e F z$, where $e>0$ is the elementary charge and $F$ is the electric field strength. This breaks the structural inversion symmetry and effectively introduces a Rashba-type SOI in the heterostructure. The VBE change across the heterostructure, which includes the effect of strain, and the bias voltage collectively shape the out-of-plane hole confinement:
\begin{equation}
\label{eq:longitudinal-confinement}
V_{\perp}^{(w)}(z) = \text{VBE}[c_w(z)] + V_\text{bias}(z),
\end{equation}
where $w$ indicates the type of well.

Meanwhile, the in-plane dot confinement is modeled as an isotropic harmonic potential:
\begin{equation}
\label{eq:lateral-confinement}
V_\parallel(x,y) = \frac{1}{2}m_\parallel\omega_\parallel^2\left(x^2 + y^2\right),
\end{equation}
where $\omega_\parallel = \frac{\hbar}{2m_\parallel \sigma_\parallel^2}$ and $m_\parallel$ is the in-plane effective mass of the hole, and $\sigma_\parallel$ is the characteristic in-plane confinement length along the $xy-\text{plane}$. In this work, we consider $\sigma_\parallel \in \left[5\,\text{nm},50\,\text{nm}\right]$. Thus, $d_w$ and $\sigma_\parallel$ determine the geometric profile of the hole wavefunction and influence the anisotropic properties of the qubit.

%%%%%%%%%%%%%%%%%%%%%%%%%%%%%%%%%%%%%%%%%%%%%%%%%%%

\section{Quantum Well Comparison}
\label{sec:well-comparison}
There are several key differences between the square well and the parabolic well. Notably, whereas the eigenenergies of a square well scale quadratically with the quantum number, the scaling for a parabolic well is linear. This allows the parabolic well to have a much denser energy spectrum, leading to stronger band mixing effects. We now discuss other differences between the two well types that are relevant to the g-factor.

\subsection{Strain}
\label{subsec:strain}
\begin{figure}
    \centering
    \includegraphics[width=\linewidth]{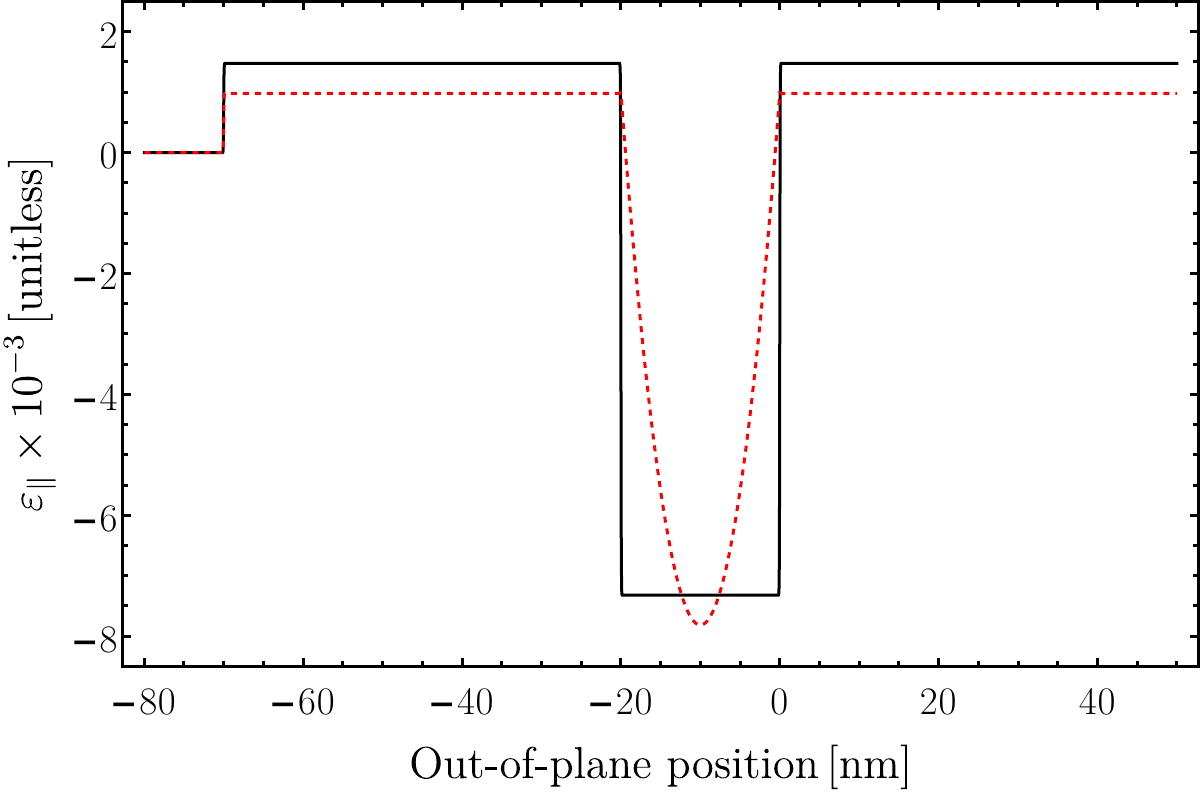}
    \caption{\justifying \small
    A plot of the strain $\varepsilon_\parallel(z)$ for both the square well (solid) and the parabolic well (dashed). The heterostructure is strain-balanced using Eq.~\eqref{eq:strain-balancing} as constraint. This prevents the formation of dislocations by matching the average lattice constant of the heterostructure to that of the buffer. Since the well has a higher Ge concentration relative to the relaxed buffer, it is compressive strained while the barrier region is tensile strained. The strain transition point is abrupt in a square well and occurs at the well-barrier interface. In contrast, the Ge concentration gradient in a parabolic well is more gradual which allows it to distribute the strain more evenly. This moves the strain transition point inside the well region instead of being at the interface.}
    \label{fig:strain}
\end{figure}

In square Ge/SiGe quantum wells, the Ge concentration step of Eq.~\eqref{eq:ge-concentration-square} at the interfaces produces strain along the plane of the well that is compressive within the well region and, if we include strain balancing, is tensile in the barriers, as shown in Fig.~\ref{fig:strain}. Although the strain is globally balanced by satisfying Eq.~\eqref{eq:strain-balancing}, the large local strain gradients at the interfaces constrain the achievable well thickness. Increasing the well thickness requires a larger concentration difference between the relaxed buffer and the bottom $\ce{Si_{0.2}Ge_{0.8}}$ barrier to maintain strain balance since more elastic energy is accumulated within the well, but this comes at the cost of a lower critical thickness~\cite{Schaffler_1997}. Consequently, growing wider wells becomes increasingly difficult without violating stability constraints unless more sophisticated strain-engineering techniques are employed.

In contrast, parabolic wells are formed by gradually changing the Ge concentration, resulting in smoother strain profiles across the heterostructure. Instead of having sharp localized strain gradients at the well-barrier interfaces, the strain is more evenly distributed, giving the parabolic well a lower average strain across the heterostructure. Unlike the square well, we see in Fig.~\ref{fig:strain} that the position where the strain changes sign (compressive to tensile or vice-versa) shifts deeper into the Ge-rich region, which avoids the abrupt stress concentration at the interfaces. This spatial smoothing of strain is particularly advantageous in cases where random strain variations can induce unwanted g-factor shifts. In addition, the reduced local strain gradient and Ge concentration lower the likelihood of strain relaxation, allowing much wider wells to be grown while satisfying the strain balancing condition with a relaxed buffer. As an example, a rough estimate of the critical thickness of a square Ge well with a $\ce{Si_{0.2}Ge_{0.8}}$ barrier is approximately ${35\,\text{nm}}$~\cite{Harrison_2016}. Meanwhile, recent experiments have demonstrated Ge/SiGe parabolic quantum wells with widths ranging approximately \SI{45}{\nano\meter} to \SI{70}{\nano\meter}~\cite{Montanari_2021,Campagna_2024}. A more careful theoretical treatment of strain effects is necessary to provide a direct and fair comparison of the critical thickness and is beyond the scope of this work.

\subsection{Qubit width}
\label{subsec:width}
Because the hole g-factor anisotropy is directly influenced by the asymmetry in the spatial profile of the wavefunction, understanding the extent of out-of-plane confinement is essential for interpreting the anisotropy trends observed in different well geometries. We begin by defining the effective physical width of the qubit as the span of one standard deviation of the position operator, which roughly captures about $68\,\%$ of the ground state wavefunction for a given direction. For example, the qubit width along the out-of-plane direction is given by $l_z = 2\sigma_z = 2\sqrt{\expval{z^2} -\expval{z}^2}$. It is important to note that $l_z \not\propto d_w$, i.e., the physical dimensions of the qubit do not directly correspond to the physical dimensions of the quantum well.

To isolate the effect of well geometry, we neglect other relevant factors, such as heavy-hole light-hole mixing and magnetic effects. Furthermore, we consider the idealized situation where both wells have infinitely high barriers to simplify the boundary conditions and allow for analytical estimates:
\begin{align}
	\label{eq:ideal-square}
    &V_{\perp}^{(\text{sq})}(z) = - e F z + 
	\begin{cases}
		0  & -d_w \leq z \leq 0,\\
		\infty & \text{otherwise},
	\end{cases}\\
    \label{eq:ideal-para}
    &V_{\perp}^{(\text{para})}(z) = - e F z + V_\text{max} \frac{(z+d_w/2)^2}{d_w^2/4},
\end{align}
where $V_\text{max} = \text{VBE}[0.8] - \text{VBE}[1]$ is the VBE offset between the 80\,\% barrier and pure Ge with strain effects included. Although these simplified models neglect finite barrier effects, they capture the essential confinement behavior and simplify our analysis. To approximate the behavior of a finite potential, we set $V_\text{max} =120\,\text{meV}$ which is consistent with empirically reported values~\cite{Schaffler_1997}. Throughout this work, we set $F = 0.5\,\text{MV/m}$ unless otherwise stated.

\begin{figure}
    \centering
    \includegraphics[width=\linewidth]{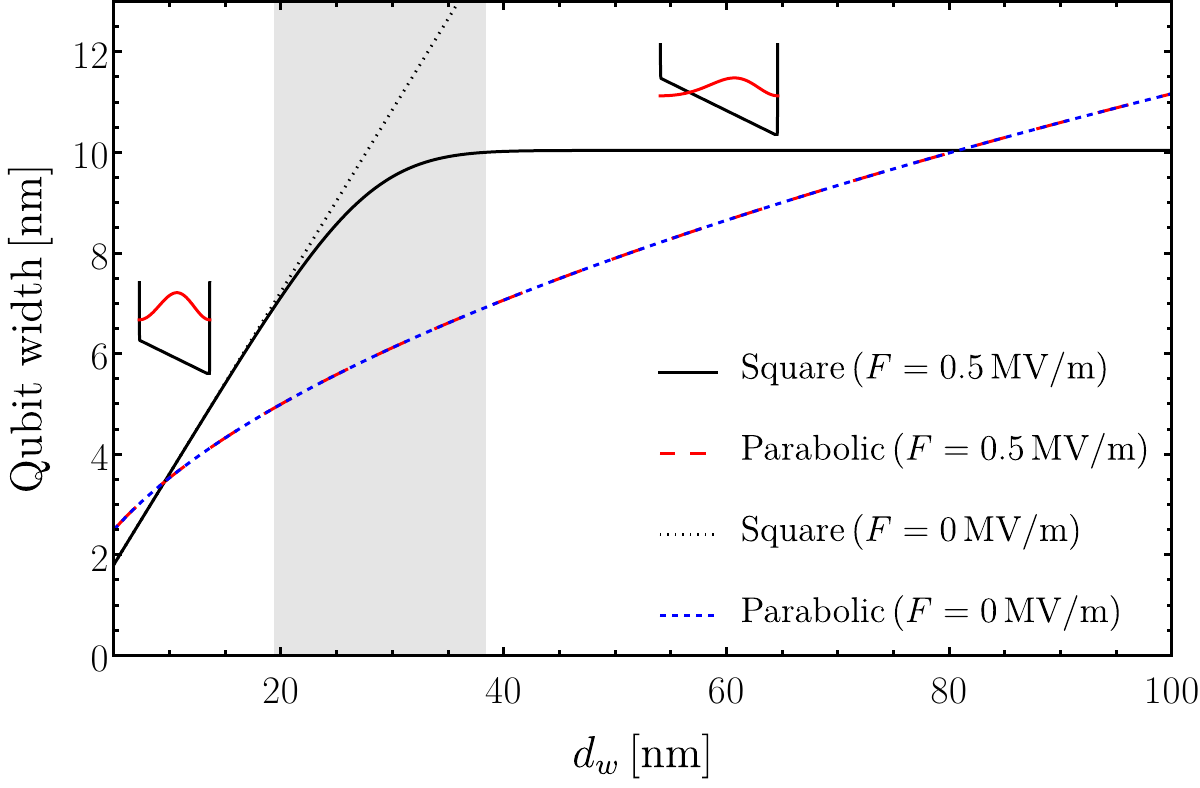}
    \caption{\justifying  \small
    A plot of the effective qubit width as a function of the quantum well width $d_w$. Here we consider the idealized quantum wells provided by Eqs.~\eqref{eq:ideal-square} and~\eqref{eq:ideal-para}. In the case of a square well, the qubit width initially increases linearly with $d_w$ as indicated in Eq.~\eqref{eq:square-scaling}. When $d_w \in \left[19.5\,\text{nm},38.3\,\text{nm}\right]$, the system enters a hybridized regime where the effect of $V_\text{bias}$ can no longer be treated perturbatively. Beyond the transition region, the system is purely dominated by $V_\text{bias}$. In this triangular well limit, the qubit width saturates to a fixed value determined by the electric field strength $F$. In contrast, the qubit width in the parabolic well is accurately estimated by Eq.~\eqref{eq:parabolic-scaling} for both values of $F$. This is because $V_\text{bias}$ only shifts the energy and position of the qubit while keeping its width the same, as we can see from Eq.~\eqref{eq:parabolic-potential-rewrite}. It is important to note, however, that the behavior of a realistic finite well potential is significantly different from this idealized case for a sufficiently large $d_w$.}
    \label{fig:scaling}
\end{figure}

We show in Fig.~\ref{fig:scaling} a plot of the qubit width for the square and parabolic wells under different electric field strengths. Our simplified well models allow us to derive analytical expressions for the limiting behaviors of $l_z$. We begin with the square well. Let $\psi_0^{(w)}$ denote the ground state of the well with energy $E_0^{(w)}$, where $w$ indicates the well type. The idealized square well has two limiting behaviors that are governed by the well thickness $d_w$: $E_0^\text{(sq)} \gg e F d_w$ and $E_0^\text{(sq)} \ll e F d_w$. In the former case, the behavior of $\psi_0^\text{(sq)}$ can be well approximated by the infinite square well ground state. Since $V_\text{bias}$ is much weaker in this regime, we can treat it as a perturbation which gives us
\begin{equation}
\label{eq:ideal-square-energy}
E_0^\text{(sq)} \xrightarrow{d_w \to 0} \frac{\pi^2\hbar^2}{2m_\perp d_w^2} + \frac{e F d_w}{2},
\end{equation}
where $m_\perp$ is the out-of-plane effective mass of the hole. In addition, the qubit length $l_z$ in this regime scales with $d_w$:
\begin{equation}
\label{eq:square-scaling}
    l_z^\text{(sq)} \xrightarrow{d_w \to 0}  \frac{d_w\sqrt{\pi^2/3-2}}{\pi},
\end{equation}
which indicates a linear relationship between $l_z$ and $d_w$ in this regime. As we further increase $d_w$, the system eventually reaches the point where $E_0^{\text{(sq)}} = e F d_w$. Beyond this point, $V_\text{bias}$ can no longer be treated perturbatively. In particular, there is a transition region that we denote by $d_w\in\left[d_\text{L}^*,d_\text{R}^*\right]$ where the system hybridizes between the two limiting behaviors. We find using Eq.~\eqref{eq:ideal-square-energy} that the transition begins at
\begin{equation}
\label{eq:characteristic-length}
    d_\text{L}^* = (2\pi^2)^{1/3}\tilde{z} = (2\pi^2)^{1/3}\left(\frac{\hbar^2}{2m_\perp e F}\right)^{1/3},
\end{equation}
where $\tilde{z}$ is the characteristic confinement length due to $V_\text{bias}$. Once we reach the other end of the transition region ($d_w = d_\text{R}^*$), $\psi_0^\text{(sq)}$ is effectively trapped in the triangular potential created by $V_\text{bias}$. To find the boundary where this occurs, we first note that the asymptotic behavior of $\psi_0^\text{(sq)}$ is given by
\begin{equation}
\label{eq:asymptotic}
    \psi_0^\text{(sq)}\sim\exp\left(-\frac{2}{3}\left(\frac{z}{\tilde{z}}\right)^{3/2}\right),
\end{equation}
which has a decay length of $\lambda = (9/4)^{1/3}\tilde{z}$. We can then define 
\begin{equation}
\label{eq:transition-end}
    d_\text{R}^* = d_\text{L}^* + 2\lambda = \left((2\pi^2)^{1/3} + 18^{1/3}\right)\tilde{z},
\end{equation}
which corresponds to a $1-\exp(-2\sqrt{2})\approx 94\,\%$ decay of $\psi_0^\text{(sq)}$ in the classically forbidden region of $V_\text{bias}$ (i.e., in the region where $E < V_\text{bias}$). In Fig.~\ref{fig:scaling}, the transition region is illustrated as the shaded region $d_w \in \left[19.5\,\text{nm},38.3\,\text{nm}\right]$. For all $d_w > d_\text{R}^*$, $l_z$ no longer scales with $d_w$ and is purely determined by the electric field strength $F$. Specifically, we have
\begin{equation}
\label{eq:triangle-scaling}
    l_z^\text{(sq)} \xrightarrow{d_w \to \infty}  \frac{4\alpha_1}{3\sqrt{5}}\tilde{z},
\end{equation}
where $-\alpha_1$ is the first zero of the Airy function $(\text{Ai}(-\alpha_1)=0)$. We see in Fig.~\ref{fig:scaling} that this results in a qubit width plateau at $l_z^\text{(sq)} = 10.04\,\text{nm}$. Thus, creating a quantum dot with sufficient lateral confinement strength is the key challenge in achieving a nearly isotropic wavefunction since current state-of-the-art lithographic techniques can only achieve confinement lengths on the order of $100\,\text{nm}$.

On the other hand, idealized parabolic wells exhibit no saturation in qubit width since $V_\text{bias}$ can be absorbed into the quadratic potential without affecting the confinement strength. To isolate the shift induced by the bias field, we can rewrite Eq.~\eqref{eq:ideal-para} as
\begin{equation}
\label{eq:parabolic-potential-rewrite}
V_\perp^\text{(para)}(z) = V_\text{max}\frac{(z-z_0)^2}{d_w^2/4} - \frac{e F}{2}\left(z_0 - \frac{d_w}{2}\right),
\end{equation}
where $z_0 = \frac{e F d_w^2}{8V_\text{max}} - \frac{d_w}{2}$. Thus, in an idealized parabolic well, $V_\text{bias}$ simply shifts the qubit position and energy while keeping the width constant. As a result, the width can be expressed as
\begin{equation}
\label{eq:parabolic-scaling}
    l_z^\text{(para)} = \left(\frac{\hbar^2 d_w^2}{2m_{\perp} V_\text{max}}\right)^{1/4},
\end{equation}
regardless of what $F$ is. We see in Fig.~\ref{fig:scaling} that, unlike in a square well, the qubit width in a parabolic well is not limited by $V_\text{bias}$. This underscores a key difference: while square wells eventually plateau in qubit width, parabolic wells continue to scale with $d_w$. 

We can make our prediction more precise by quantifying the in-plane qubit dimension $l_\parallel$. Due to our choice of parametrization, we simply have $l_\parallel = 2\sigma_\parallel$. Given the range of $\sigma_\parallel$ that we consider in this work, this corresponds to in-plane qubit dimension ranging between \SI{10}{\nano\meter} to \SI{100}{\nano\meter}. Since $l_z \lessapprox10\,\text{nm}$ in both well types for the range of $d_w$ we considered, we predict that the characteristic in-plane confinement length necessary to achieve g-factor isotropy should be $\sigma_\parallel \approx 5\,\text{nm}$. If we further consider the effect of band mixing, this estimate will shift slightly due to the differences in hole effective masses from different bands.

It is crucial to note that for a realistic, finite parabolic well, other effects will determine the limiting behavior of $l_z^\text{(para)}$. Importantly, as we increase $d_w$, the shift $z_0$ pushes the center of the parabola into the barrier region, which can significantly alter the well shape and its properties. This effect is also influenced by the bias strength $F$, with larger values causing even more distortion. In this work, we only consider parameters that yield a well-defined finite parabolic well shape.

These results demonstrate that square wells fundamentally limit the vertical extent of the wavefunction due to its interaction with $V_\text{bias}$, while parabolic wells allow for tunable, extended confinement geometries. This distinction is crucial for engineering more isotropic g-factors in real devices.

%%%%%%%%%%%%%%%%%%%%%%%%%%%%%%%%%%%%%%%%%%%%%%%%%%%

\section{Hole g-factor calculation}
\label{sec:g-factor}

Having analyzed the qualitative behavior of quantum well confinement in idealized models, we now present our numerical framework for calculating the hole spin qubit g-factor in realistic Ge/SiGe quantum dots to quantify the effect of dot geometry on the anisotropy. We model the total qubit Hamiltonian as
\begin{equation}
\label{eq:total-Hamiltonian}
H  = H_{\text{BF}} + V_\text{dot}^\text{(w)}  + H_\text{Z},
\end{equation}
where $H_{\text{BF}}$ is the Burt-Foreman Hamiltonian which describes the kinetic energy of valence holes as well as band mixing, $V_\text{dot}^{(w)} $ is the dot confinement potential which accounts for strain through the Bir-Pikus Hamiltonian, and $H_\text{Z}$ is the Zeeman Hamiltonian. We first discuss the details of each Hamiltonian term in Eq.~\eqref{eq:total-Hamiltonian}.

\subsection{Qubit Hamiltonian}
\label{subsec:hamiltonian}
The $H_\text{BF}$ describes the dynamics of valence hole states in a quantum dot. Many theoretical works rely on the Luttinger-Kohn Hamiltonian for this task; however, $H_\text{BF}$ is better suited for this task since it naturally enforces proper operator ordering without relying on ad hoc symmetrization procedures~\cite{Willatzen_2009}. Starting from exact envelope function theory, second-order perturbation theory is used to project $H_{\text{BF}}$ onto the valence spin-3/2 subspace spanned by the eigenstates of $\hat{J}_z$ angular momentum operator, commonly known as the heavy hole (HH), $\ket{\pm\frac{3}{2}}$, and light hole (LH), $\ket{\pm\frac{1}{2}}$, states:
\begin{equation}
\label{eq:BF-Hamiltonian}
H_{\text{BF}} = 
\quad 
\begin{blockarray}{cccc}
\ket{+\frac{3}{2}} & \ket{+\frac{1}{2}} & \ket{-\frac{1}{2}} & \ket{-\frac{3}{2}} \\
\begin{block}{(cccc)}
 P + Q & S_- & -R & 0 \\
 S_-^\dagger & P - Q & -C & R \\
 -R^\dagger & -C^\dagger & P - Q & S_+^\dagger \\
 0 & R^\dagger & S_+ & P + Q \\
\end{block}
\end{blockarray}
\quad,
\end{equation}
\begin{gather}
    P = \frac{\hbar^2}{2 m_0}\left(\gamma_1 k_\parallel^2 + k_z \gamma_1 k_z\right),\\
    Q = \frac{\hbar^2}{2 m_0}\left(\gamma_2 k_\parallel^2 - 2k_z \gamma_2 k_z\right),\\
    S_\pm = \sqrt{3}\frac{\hbar^2}{2 m_0}\left( k_\pm \left(\gamma_3 + \chi\right) k_z + k_z\left(\gamma_3 - \chi\right)k_\pm \right),\\
    R = -\frac{\sqrt{3}}{2}\frac{\hbar^2}{2 m_0}\left( k_+ \left(\gamma_2 - \gamma_3\right) k_+ + k_- \left(\gamma_2 + \gamma_3\right) k_-\right),\\
    C = \frac{\hbar^2}{m_0}\left( k_- \chi k_z - k_z \chi k_- \right),
\end{gather}
where $m_0$ is the bare electron mass, $k_\pm = k_x \pm i \hat{k}_y$, and each $k_a = -i \partial_i$ with $i = x,y, z$. We reiterate that the out-of-plane, or more specifically the [001] growth direction, is the orientation of the $z$-axis while the dot plane is spanned by the $xy-\text{plane}$. The variables $\gamma_i = \gamma_i[c_\text{sq/para}]$ denote the Luttinger parameters and the asymmetry parameter $\chi = \left( 2\gamma_2 + 3\gamma_3 - \gamma_1 - 1\right)/3$. We also emphasize that, unless stated otherwise, all material parameters are composition-dependent and are calculated by interpolating between their corresponding Ge and Si parameter values as described by the composition profile of the heterostructure in Eqs.~\eqref{eq:ge-concentration-square} and ~\eqref{eq:ge-concentration-para}. For notational convenience, we will drop the explicit compositional dependence when referring to material parameters.

We can identify from Eq.~\eqref{eq:BF-Hamiltonian} the following effective mass parameters that are associated with the in-plane and out-of-plane dynamics of the HH and LH states:
\begin{gather*}
	m_{\parallel,\text{HH}} = \frac{m_0}{\gamma_1 + \gamma_2},
	m_{\perp,\text{HH}} = \frac{m_0}{\gamma_1 - 2\gamma_2},\\
	m_{\parallel,\text{LH}} = \frac{m_0}{\gamma_1 - \gamma_2},
	m_{\perp,\text{LH}} = \frac{m_0}{\gamma_1 + 2\gamma_2}.
\end{gather*}
If we take pure Ge well as an example, we have $m_{\parallel,\text{HH}} \approx 0.06m_0$, $m_{\perp,\text{HH}} \approx 0.20m_0$, $m_{\parallel,\text{LH}} \approx 0.11m_0$, and $m_{\perp,\text{LH}} \approx 0.05m_0$. We see immediately that there is significant anisotropy in the material parameters that is intrinsic to the heterostructure. The severity of this anisotropy plays a crucial role in how the confinement strength of the dot may be engineered to produce more isotropic g-factors. We model dot confinement using Eq.~\eqref{eq:lateral-confinement} and a modified version of Eq.~\eqref{eq:longitudinal-confinement} that accounts for different hole types:
\begin{equation}
    V_\text{dot}^{(w)} = \text{diag}\left(V_{\perp,j}^{(w)}(z) + V_{\parallel,j}(x,y)\right),
\end{equation}
with $j \in \left\lbrace\text{HH},\text{LH}\right\rbrace$ denoting the type of hole. 

The effects of strain on different hole states is included in $V_{\perp,j}^{(w)}(z)$ by combining the reported VBE in Ref.~\cite{Schaffler_1997}, which includes hydrostatic strain effects, and uniaxial strain correction through the Bir-Pikus Hamiltonian. Throughout our work, we only consider biaxial strain, which means that off-diagonal shearing elements in the strain tensor are neglected. This leaves us with a purely diagonal strain tensor with components $\epsilon_{xx},\epsilon_{yy},$ and $\epsilon_{zz}$. The first two variables can be described using Eq.~\eqref{eq:strain} (i.e., $\epsilon_{xx} = \epsilon_{yy} = \epsilon[c(z)]$) while $\epsilon_{zz} = -2\frac{C_{12}}{C_{11}}\epsilon[c(z)]$, where the factor represents Poisson's ratio for cubic semiconductors. This gives us the following strain or Bir-Pikus Hamiltonian for:
\begin{equation}
\label{eq:bir-pikus}
    H_\text{BP}(z) = \text{diag}\left(Q_\epsilon,-Q_\epsilon,-Q_\epsilon,Q_\epsilon\right),
\end{equation}
where $Q_\epsilon = - \frac{b_v}{2}\left(\epsilon_{xx} + \epsilon_{yy} - 2\epsilon_{zz}\right)$ and $b_v$ is a deformation potential. Thus, the effect of strain on the HH and LH states can be modeled by modifying Eq.~\eqref{eq:longitudinal-confinement} into
\begin{equation}
\label{eq:longitudinal-confinement-modified}
V_{\perp}^{(w)}(z) = H_\text{BP}(z) + \left(\text{VBE}(z) + V_\text{bias}(z)\right) \text{Id},
\end{equation}
where $\text{Id}$ is the identity operator.

Next, we also modify $V_{\parallel}(x,y)$ by multiplying it by $\text{Id}$. We note that since the ground state of the system has a mostly HH-like character due to the dominance of compressive strain in the well, we choose the confinement strength so that $\omega = \frac{\hbar}{2m_{\parallel,\text{HH}}l_\parallel^2}$.%, where $l_\parallel = l_x = l_y$.

Finally, magnetic interactions are perturbatively included through the Zeeman Hamiltonian:
\begin{equation}
\label{eq:zeeman}
    H_\text{Z} = -2\mu_B \kappa \mathbf{J}\cdot\mathbf{B} - 2\mu_B q\mathbf{\mathcal{J}}\cdot\mathbf{B},
\end{equation}
where $\mu_B$ is the Bohr magneton and $\kappa$ (q) are the (an)isotropitc contribution of the g-factor. The magnetic field $\mathbf{B} = (B_x, B_y, B_z)$ is generated by a vector potential, which we choose to be $\mathbf{A} = (x B_z/2 - z B_x, -y B_z/2 + z B_y, 0)$ for convenience. In addition, the magnetic field couples to the orbital degree of freedom, which can be captured by modifying Eq.~\eqref{eq:BF-Hamiltonian} according to the Peierls substitution: $\mathbf{k} \rightarrow \mathbf{k} - \frac{e}{\hbar} \mathbf{A}$, where $e$ is the elementary charge. In this work, we set $B = 0.1\,\text{T}$.

\subsection{Effective hole g-factor}
\label{subsec:g-factor}

With the dot Hamiltonian in place, we now compute the effective hole g-factor. First, we project Eq.~\eqref{eq:total-Hamiltonian} further onto a basis of hole subbands that is spanned by the diagonal elements of Eq.~\eqref{eq:total-Hamiltonian} in the absence of magnetic field (i.e., we neglect $H_\text{Z}$ and all off-diagonal terms in $H_\text{BF}$). This gives us two separable equations:
\begin{gather}
    \left[-\frac{\hbar^2}{2}\partial_z\frac{1}{m_{\perp,i}(z)}\partial_z + V_\perp^{(w)}(z)\right]\eta(z) = E_\perp \eta(z)
    \label{eq:longitudinal-basis-H} \\
    \left[-\frac{\hbar^2}{2m_{\parallel,i}}\left(\partial_x^2 + \partial_y^2\right) + \frac{1}{2}m_{\parallel,i}\omega^2\left(x^2+y^2\right)\right]\nonumber\\
    \times \nu(x,y) = E_\parallel \nu(x,y),
    \label{eq:transverse-basis-H}
\end{gather}
where $i\in\left\lbrace\text{HH,LH}\right\rbrace$. The states denoted as $\eta(z)$ are essentially 1-dimensional quantum well eigenstates, whereas $\nu(x,y)$ are the eigenstates of a 2-dimensional isotropic quantum harmonic oscillator. We solve Eqs.~\eqref{eq:longitudinal-basis-H} and \eqref{eq:transverse-basis-H} to obtain the HH and LH eigenstates, which form the basis
onto which Eq.~\eqref{eq:total-Hamiltonian} is projected: $\psi_i(x,y,z) = \nu_i(x,y)\eta_i(z)$. We then numerically diagonalize the resulting matrix and extract the two lowest eigenenergies, $E_0$ and $E_1$, which define our qubit subspace. The effective hole g-factor is then calculated as
\begin{equation}
\label{eq:g-factor}
    g \equiv (E_1-E_0)/\mu_B B.
\end{equation}
The in-plane $(g_\parallel)$ and out-of-plane $(g_\perp)$ g-factors are defined as the Zeeman splittings under $B_z = 0$ and $B_x = B_y = 0$, respectively.

We note that for analytical simplicity, the effective mass in Eq.~\eqref{eq:transverse-basis-H} is fixed to its corresponding value in pure Ge. This approximation assumes that the out-of-plane dynamics is decoupled from the in-plane dynamics, which is an assumption that is valid for many devices where $l_\parallel \gg l_z$. Although we do not go over the details of how to tackle this more carefully, we note that more basis states would be required to accurately represent the true eigenstates as we explore cases where $l_\parallel \approx l_z$. Nevertheless, we also use atomistic tight-binding models to qualitatively verify the reduction of out-of-plane g-factors ($g_\perp$) and the g-factor anisotropy ($g_\perp/g_\parallel$) in regions where where $l_\parallel \gg l_z$ (see Appendix~\ref{app:tb}).

\begin{widetext}
	\onecolumngrid
\begin{figure*}[ht]
	\centering
	\subfloat[Square $g_\parallel$]{
		\includegraphics[width=.31\linewidth]{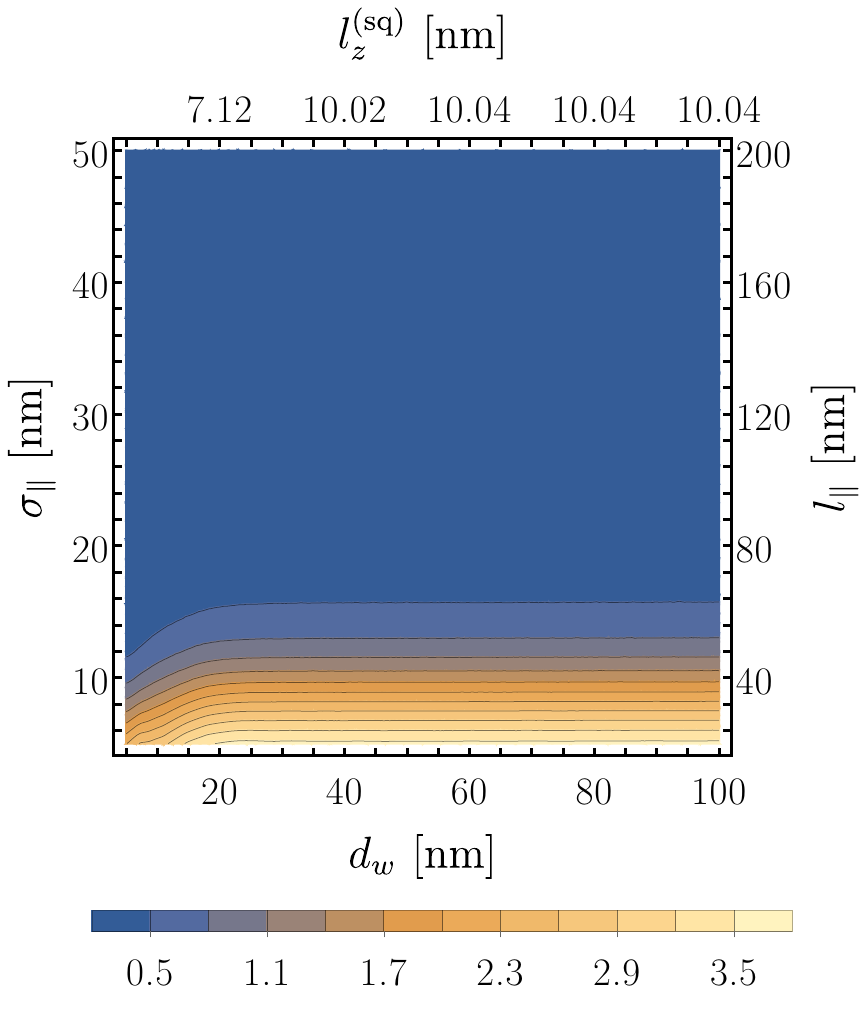}
		\label{subfig:contour-capped-square-g_para}
	}
	\subfloat[Square $g_\perp$]{
		\includegraphics[width=.31\linewidth]{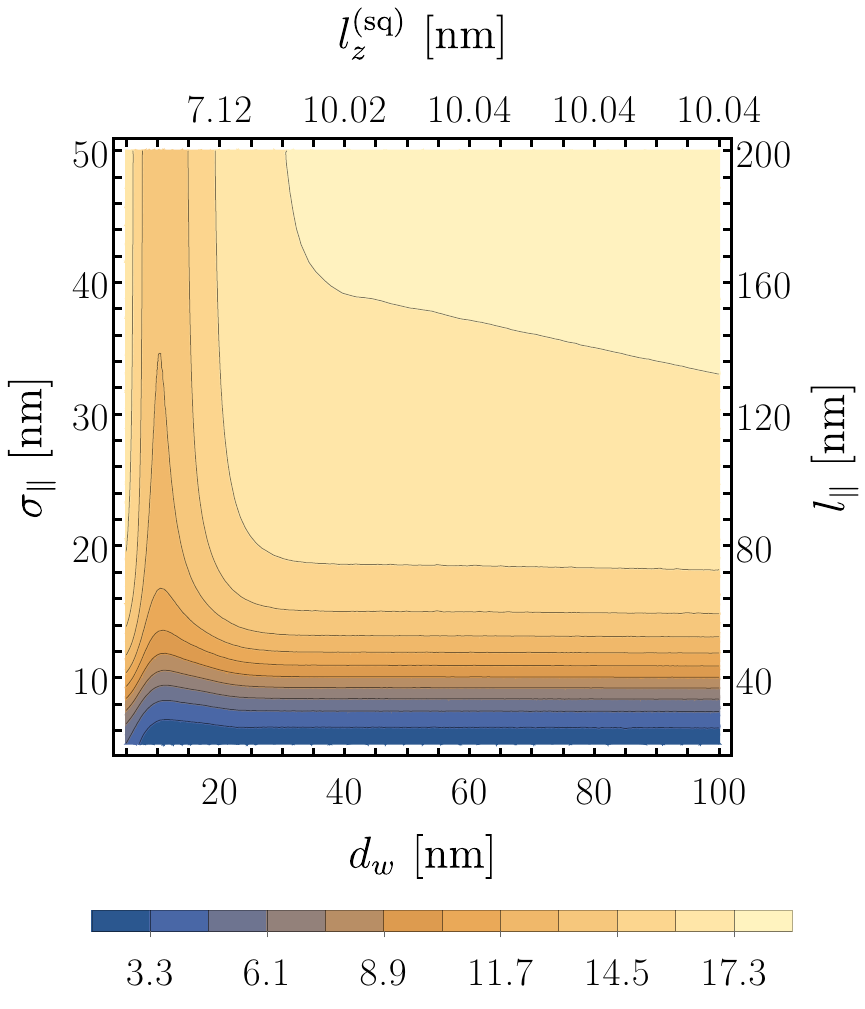}
		\label{subfig:contour-capped-square-g_perp}
	}
	\subfloat[Square $g_\perp/g_\parallel$]{
		\includegraphics[width=.31\linewidth]{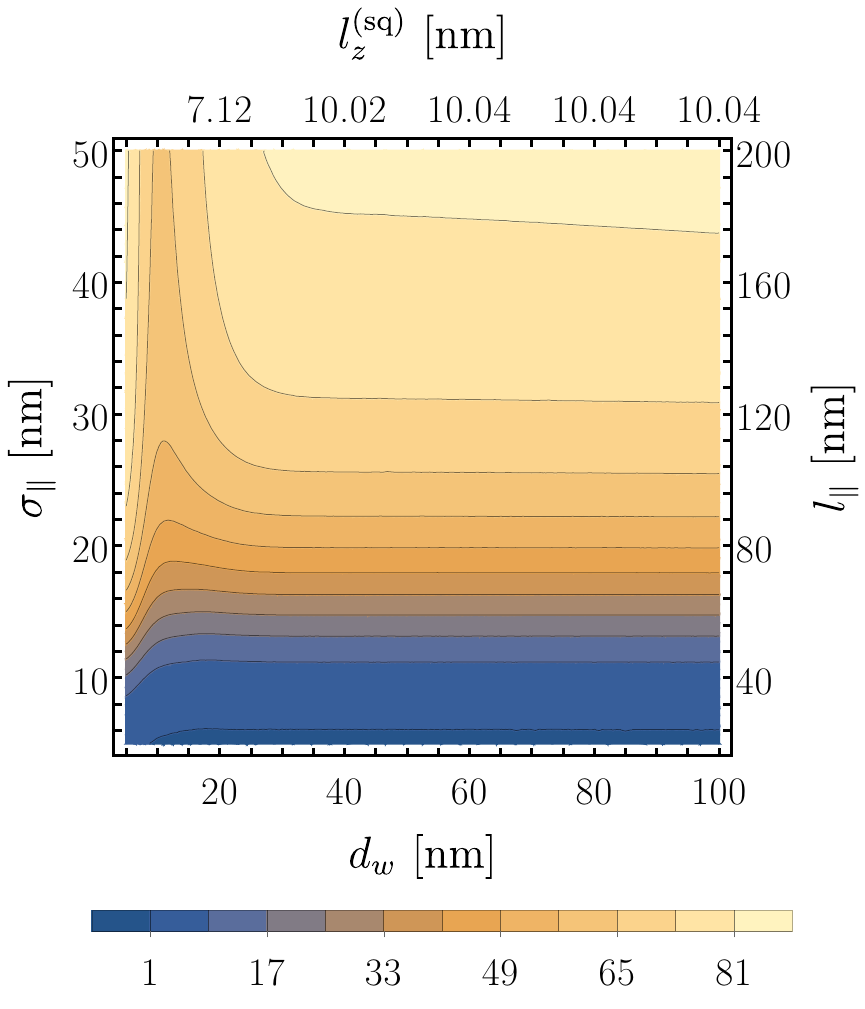}
		\label{subfig:contour-capped-square-g_ratio}
	}
	\\
	\subfloat[Parabolic $g_\parallel$]{
		\includegraphics[width=.31\linewidth]{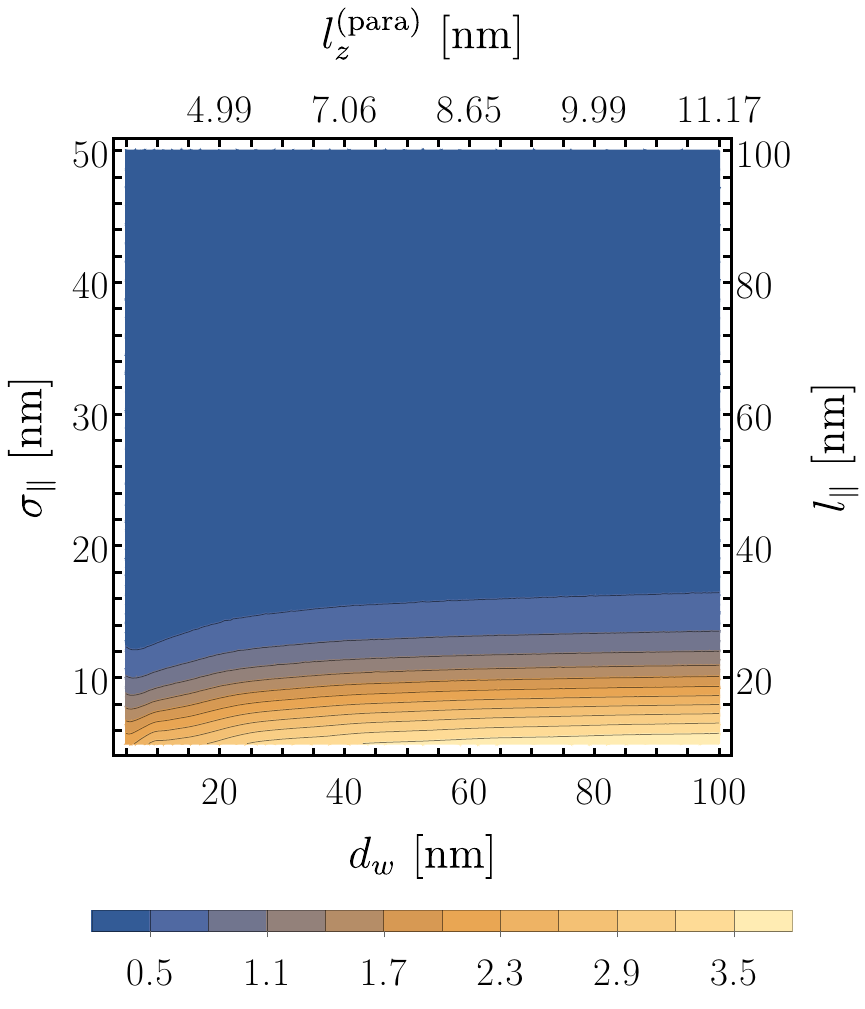}
		\label{subfig:contour-capped-parabolic-g_para}
	}
	\subfloat[Parabolic $g_\perp$]{
		\includegraphics[width=.31\linewidth]{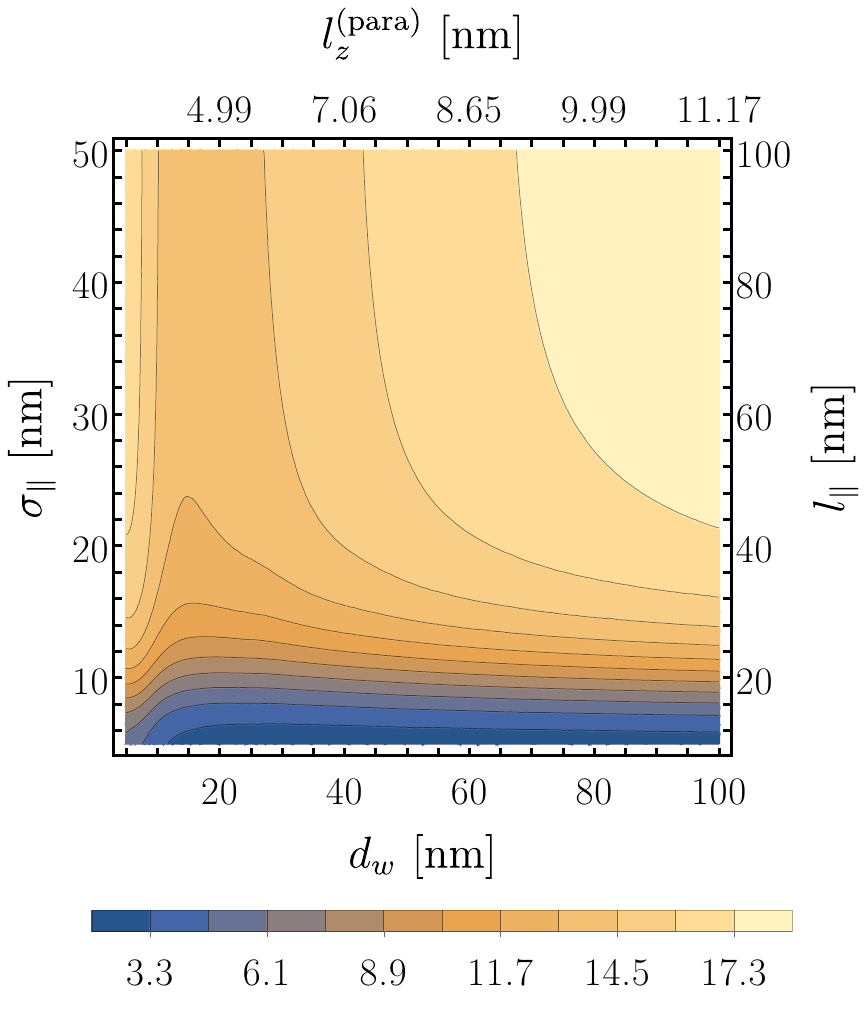}
		\label{subfig:contour-capped-parabolic-g_perp}
	}
	\subfloat[Parabolic $g_\perp/g_\parallel$]{
		\includegraphics[width=.31\linewidth]{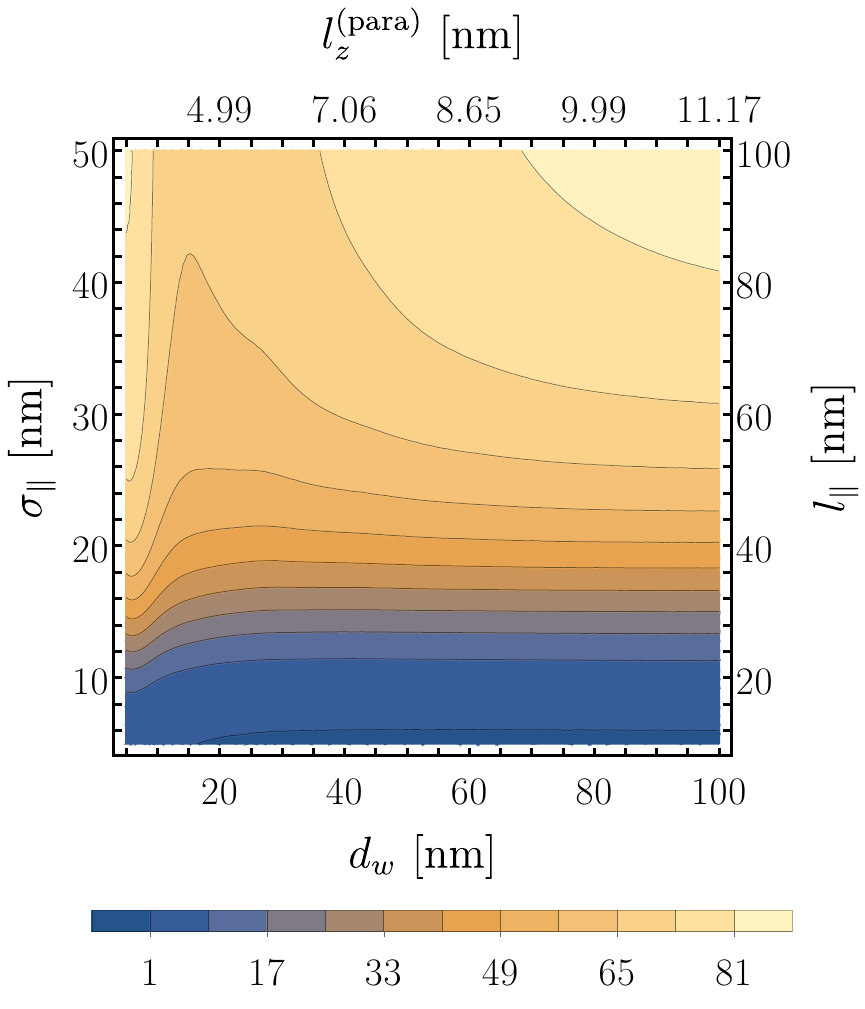}
		\label{subfig:contour-capped-parabolic-g_ratio}
	}
	\caption{\justifying \small
    Contour plots of effective g-factor and anisotropies for the square and parabolic quantum well. The square well g-factor anisotropy plateaus for large $d_w$ because the system becomes dominated by the triangular potential $V=-e F z$. In contrast, the parabolic well continuously changes with $d_w$, demonstrating the potential for qubit engineering. Importantly, we verify our intuition in both well types that increasing the lateral confinement strength to make $l_z \approx l_\parallel$ reduces the g-factor anisotropy.
    }
    \label{fig:contour-capped}
\end{figure*}
\end{widetext}

\subsection{Numerical results}
\label{subsec:results}
We now present the numerical results of our g-factor calculations. Our qubit Hamiltonian is generated by projecting Eq.~\eqref{eq:total-Hamiltonian} onto the lowest 20 HH and 20 LH well eigenstates. We further project onto 146 2D harmonic oscillator eigenstates with $|{L_z}|\leq 5\hbar$. All sharp features or discontinuities in the material parameters and potentials are smoothed using a sigmoid function.

We show in Fig.~\ref{fig:contour-capped} the contour plots of the g-factors $g_\perp$, $g_\parallel$, and the anisotropy $g_\perp/g_\parallel$ as functions of the well depth $d_w$ and the effective in-plane confinement length $\sigma_\parallel$. For the moment, we have excluded the effects of strain balancing, which means $c_0 = 0.8$, the same as the barrier concentration. We included in this figure the corresponding $l_\parallel$ and $l_z$ predicted from Sec.~\ref{subsec:width} for comparison. Our numerical results show that the application of stronger lateral confinement reduces the anisotropy in both well types. In particular, we observe that the g-factor in either well type becomes nearly isotropic (i.e., $g_\perp/g_\parallel = 1$) when $l_z \approx \sigma_\parallel$. Unfortunately, such a strong lateral confinement is not feasible with the current state-of-the-art. As an example, let us suppose that we have a typical planar Ge quantum dot with $d_w = 20\,\text{nm}$ and $\sigma_\parallel = 50\,\text{nm}$. Even an order of magnitude reduction in the g-factor anisotropy would require a lateral confinement of roughly $\sigma_\parallel = 10\,\text{nm}$ for either well type. Nevertheless, these results corroborate our theoretical prediction and, more importantly, they support our intuition that isotropic wavefunctions give rise to isotropic g-factors.

\begin{figure}
    \centering
    \subfloat[HH-LH mixing]{
    \includegraphics[width=\linewidth]{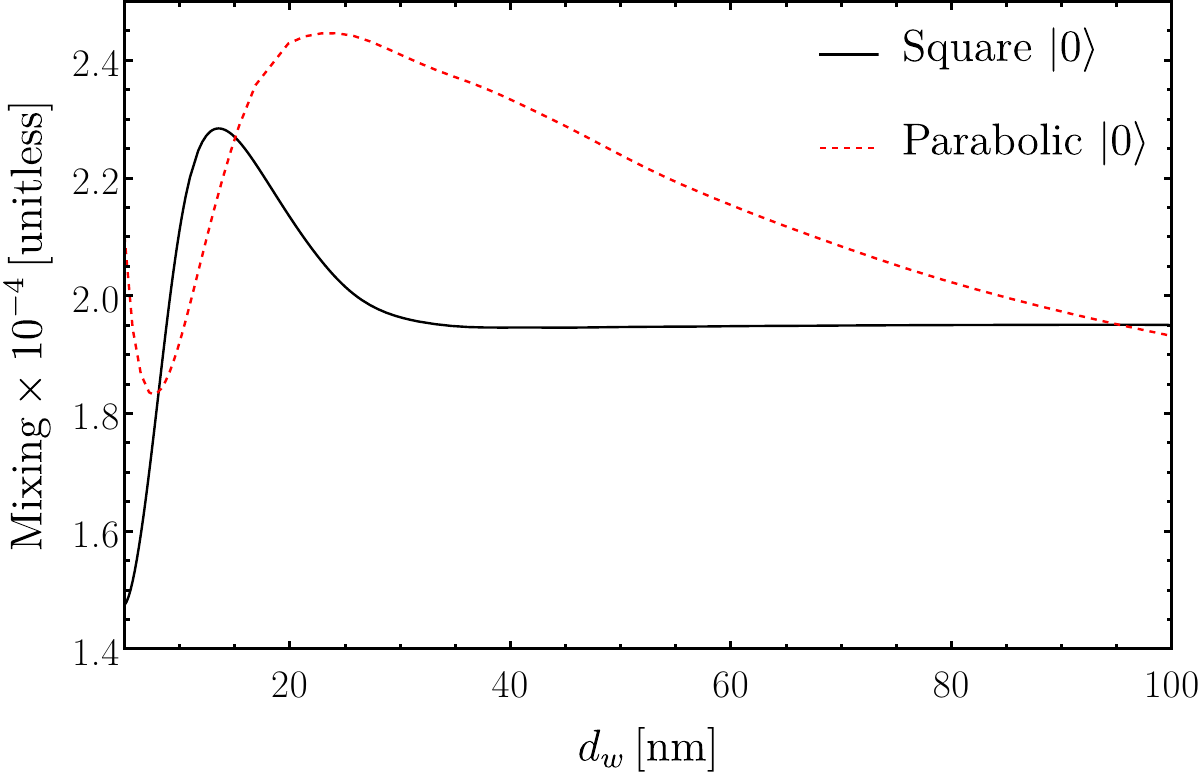}
    }\\
    \subfloat[$g_\perp$]{
    \includegraphics[width=\linewidth]{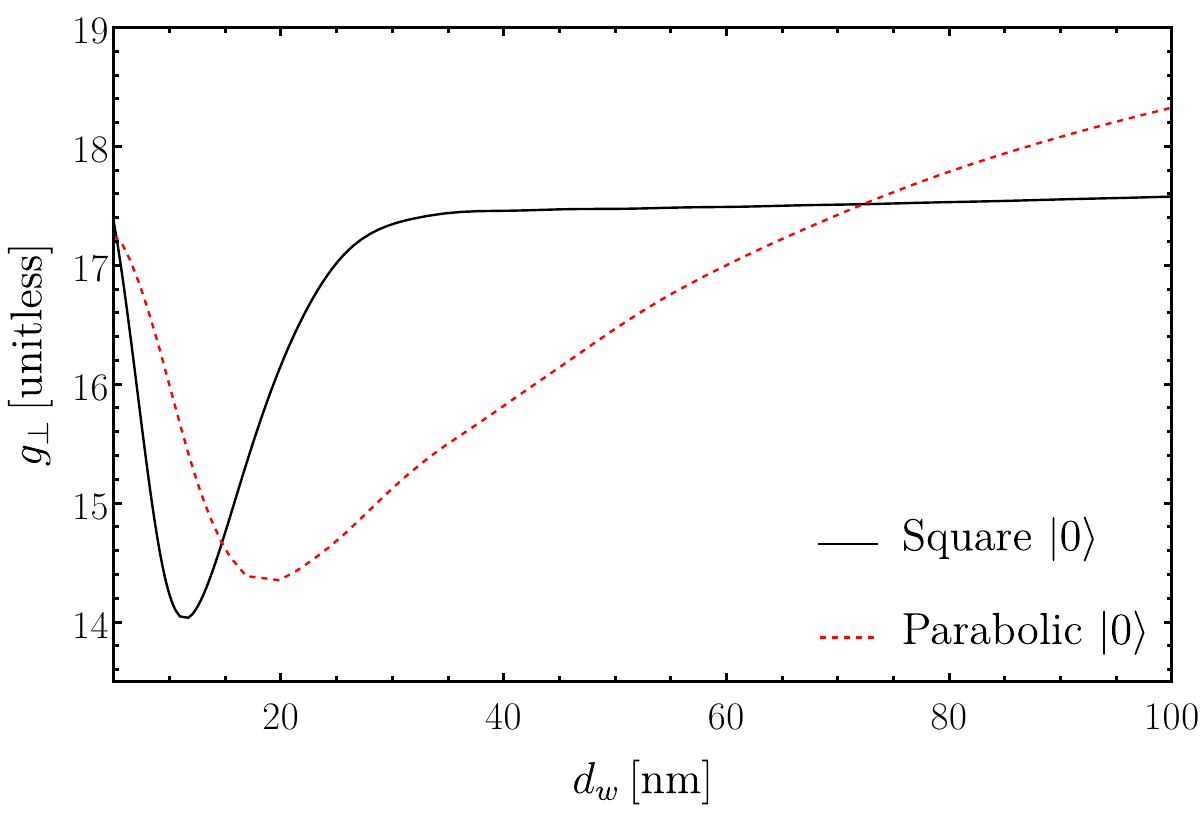}
    }
    \caption{\justifying \small
    A plot of (a) HH-LH mixing and (b) $g_\perp$ as a function of the well thickness $d_w$ when $\sigma_\parallel = 50\,\text{nm}$. We observe in both well types that an increase in mixing decreases $g_\perp$ and vice-versa. For the square well, we find that the mixing plateaus in the region predicted in Sec.~\ref{subsec:width}. In contrast, the mixing for the parabolic well initially increases for thin wells but decreases eventually. This is because of the parabolic nature of the potential and the difference in the HH and LH effective masses. As $d_w$ increases, the center of the parabolic well shifts according to Eq.~\eqref{eq:parabolic-potential-rewrite}. The difference in effective masses makes the HH and LH states shift differently, resulting in less wavefunction overlap and, therefore, HH-LH coupling matrix elements.}
    \label{fig:mixing}
\end{figure}

Next, we discuss the dependence of the g-factor on $d_w$. We observe a valley-like drop in $g_\perp$ for both well types when the quantum well is sufficiently thin ($d_w < 20\,\text{nm}$). We attribute this feature to the number of bound LH eigenstates within the well. Initially, at low $d_w$, only one bound $\eta_\text{LH}(z)$ state is allowed inside while the rest of the LH well eigenstates are delocalized. As $d_w$ increases, a second bound eigenstate becomes allowed which results in a significant change in g-factor behavior. Although more bound state become available if we widen $d_w$ significantly, the change in g-factor is much less due to the energy difference between the ground HH and the excited LH states. 

As $d_w$ increases further, we reach the asymptotic limit for the square well where $l_z^\text{(sq)}$ becomes practically constant as predicted in Sec.~\ref{subsec:width}. Indeed, we see this constant-$g_\perp$ behavior manifest within the transition region $[19.5\,\text{nm},38.3\,\text{nm}]$ of the square well. In contrast, the parabolic well continuously changes with $d_w$, allowing for more $g_\perp$ tunability. In particular, when $\sigma_\parallel \gtrapprox 7\,\text{nm}$, we find that the $g_\perp$ initially decreases with $d_w$ (as a result of the second LH state becoming available) and eventually begins increasing with $d_w$. However, when the in-plane confinement is very strong ($\sigma_\parallel < 10\,\text{nm}$), $g_\perp$ becomes monotonically decreasing with $d_w$. Unfortunately, our current theoretical framework is not ideal to analyze the properties of this parameter regime since the separability of $\eta(z)$ and $\nu(x,y)$ would no longer valid. Finally, $g_\parallel$ is largely unaffected by changes in $d_w$ in both well types, except for when $d_w$ and $\sigma_\parallel$ are sufficiently small that $\eta-\nu$ coupling becomes relevant. 

The general behavior of the g-factor can also be predicted from the HH-LH mixing. We formally quantify HH-LH mixing as the projection of the ground eigenstate of Eq.~\eqref{eq:total-Hamiltonian} onto the $\ket{\pm\frac{1}{2}}$ subspace: $\text{proj}_{\pm\frac{1}{2}}\psi_0 = |\bra{\frac{1}{2}}\ket{\psi_0}|^2+|\bra{-\frac{1}{2}}\ket{\psi_0}|^2$. We show in Fig.~\ref{fig:mixing} the HH-LH mixing and $g_\perp$ plots as a function of $d_w$ for $\sigma_\parallel = 50\,\text{nm}$. We immediately see that there is a direct correspondence between mixing and $g_\perp$: increasing HH-LH mixing reduces $g_\perp$ and vice-versa. This is a well-established relationship and has been noted before in the literature~\cite{Scappucci_2021}. We can use this as a new framework to explain the increase in $g_\perp$ for a parabolic well, which is not something that can be explained by our theory in Sec.~\ref{subsec:width}, despite the decrease in mixing. There are two mechanisms at play: the difference between the HH and LH VBE and the parabolic nature of the potential. Recall from Eq.~\eqref{eq:parabolic-potential-rewrite} that the potential shifts by $z_0$, which is a function of $d_w$, $F$, and $V_\text{max}$. As $d_w$ increases, the parabolic well is shifted towards the top $\ce{Si_{0.2}Ge_{0.8}}$ barrier. When $F = 0$, the HH and LH wells have the same shift. However, when $F$ is nonzero, the difference between $V_\text{max,HH}$ and $V_\text{max,LH}$ produces unequal HH and LH well shifts. The well shift discrepancy grows with increasing $d_w$, resulting in less HH and LH wavefunction overlap and, consequently, less mixing. In contrast, the mixing behavior in a square well is primarily determined by its shape. Whereas parabolic wells are defined by their fixed vertex and boundary Ge concentration, the fixed Ge concentration in a square well results in the asymptotic behavior. Thus, the mixing behavior in parabolic wells is fundamentally different from that of square wells.

\subsection{Strain balancing}
\label{subsec:strain-balancing}

We finally consider the effect of strain balancing, focusing particularly on parabolic quantum wells. Strain balancing reduces the overall strain within the system, making its characteristics more bulk-like. As a result, the HH-LH splitting is reduced and the band mixing increases~\cite{Mauro_2025_2,Costa_2025_arxiv}. Fig.~\ref{fig:strain-balance} shows the modified $g_\perp$ and mixing behavior when the strain is balanced through Eq.~\eqref{eq:strain-balancing}. Overall, $g_\perp$ decreases relative to the non-strain-balanced case, which we attribute to this enhanced mixing.

However, the previously noted correspondence between the mixing and $g_\perp$ trends does not generally hold once strain balancing is introduced. As seen in Figs.~\ref{subfig:balanced-gperp} and \ref{subfig:balanced-mixing}, this correspondence breaks down for well widths exceeding roughly $20\,\text{nm}$. This behavior can be better understood by examining how strain balancing affects the well potential. As $d_w$ increases, the depth of the potential well changes by less than $1\,\text{meV}$ over the range of values we considered, indicating that the qualitative behavior of the well eigenstates is largely unchanged. Consequently, the off-diagonal matrix elements coupling the HH and LH states in Eq.~\eqref{eq:BF-Hamiltonian}, of the form $\mel{\pm\frac{3}{2}}{S_\pm,R}{\pm\frac{1}{2}}$, also remain essentially unaffected and are comparable in value to those of the non-strain-balanced case.

In contrast, Fig.~\ref{subfig:balanced-offset} shows that the HH-LH offset decreases significantly with increasing $d_w$. According to perturbation theory, this reduced HH-LH energy separation leads to stronger mixing (Fig.~\ref{subfig:balanced-mixing}), which in turn lowers $g_\perp$ (Fig.~\ref{subfig:balanced-gperp}). Meanwhile, the HH-LH offset is fixed in the non-strain-balanced case discussed in Sec.~\ref{subsec:results}. Thus, strain balancing effectively introduces the HH-LH offset as a tunable parameter, adding another layer of complexity to the resulting $g_\perp$ behavior.

\begin{widetext}
\onecolumngrid
    \begin{figure*}
    \subfloat[Strain-balanced $g_\perp$]{
    \includegraphics[width=.325\linewidth]{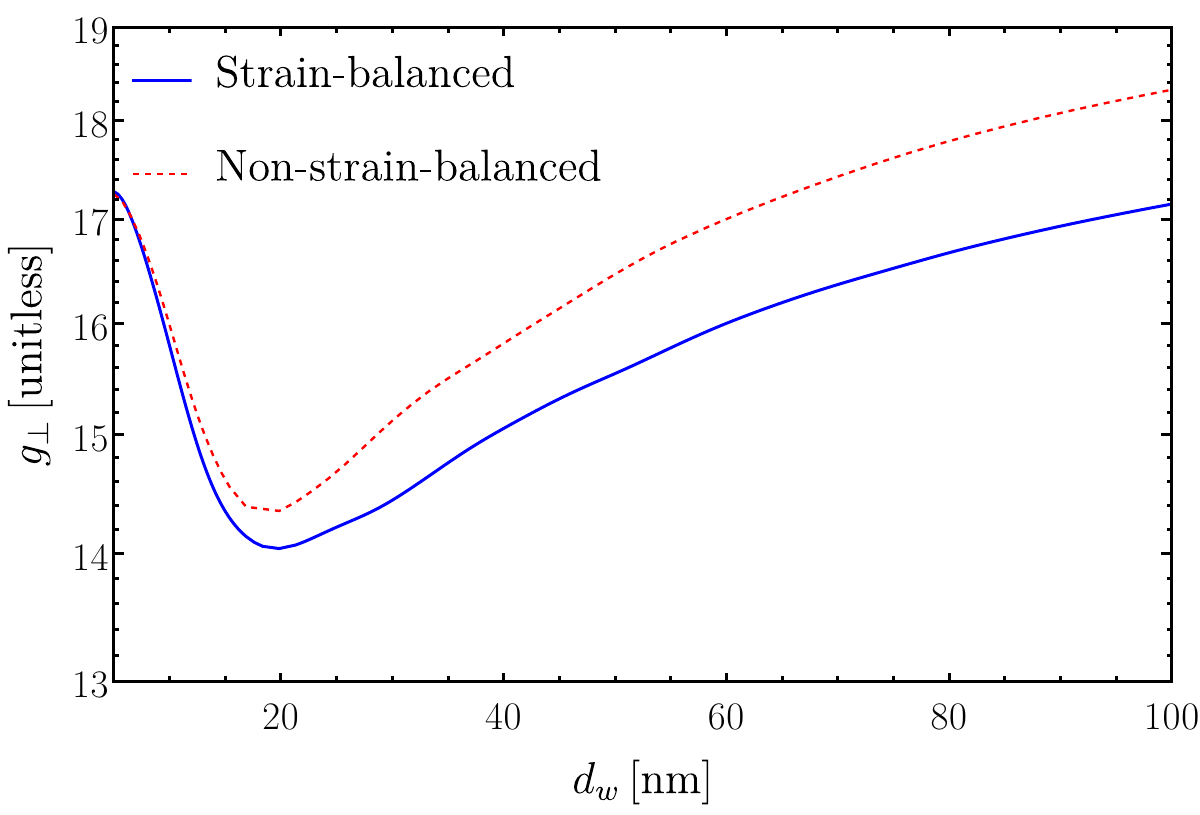}
    \label{subfig:balanced-gperp}
    }
    \subfloat[Strain-balanced mixing]{
    \includegraphics[width=.325\linewidth]{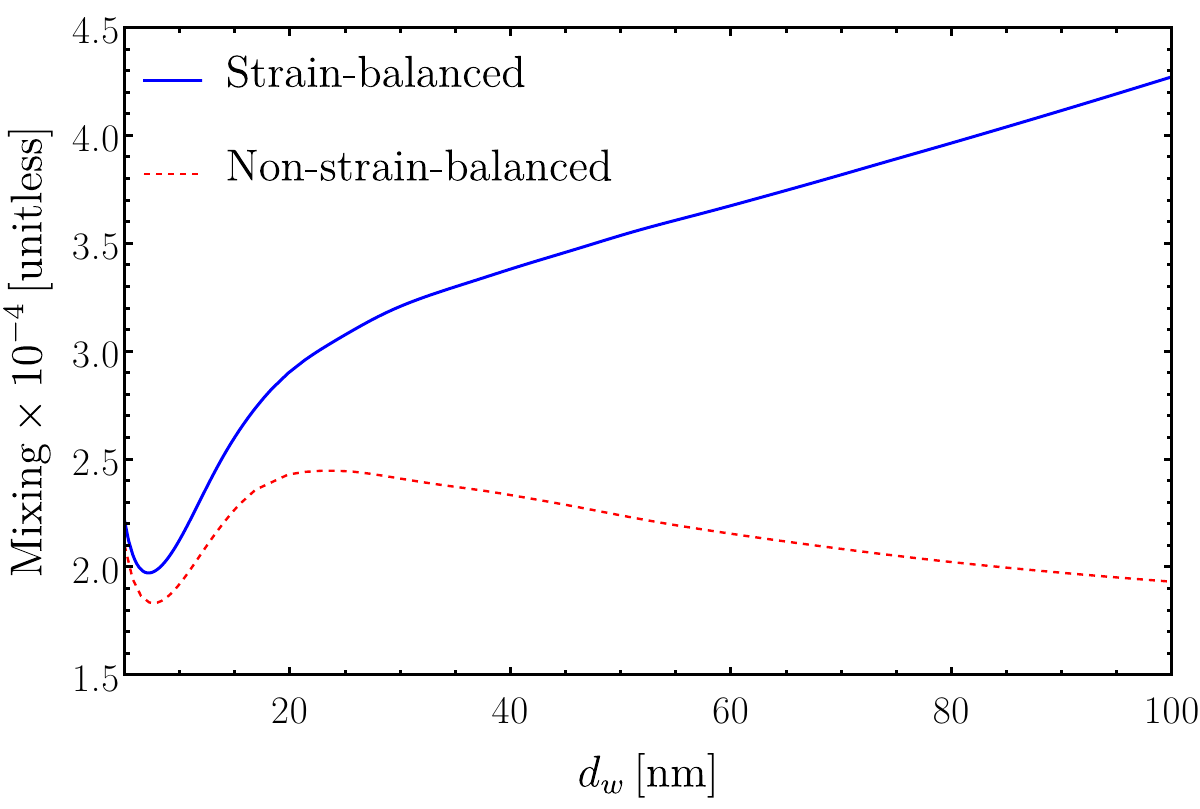}
    \label{subfig:balanced-mixing}
    }
    \subfloat[Strain-balanced HH-LH offset]{
    \includegraphics[width=.325\linewidth]{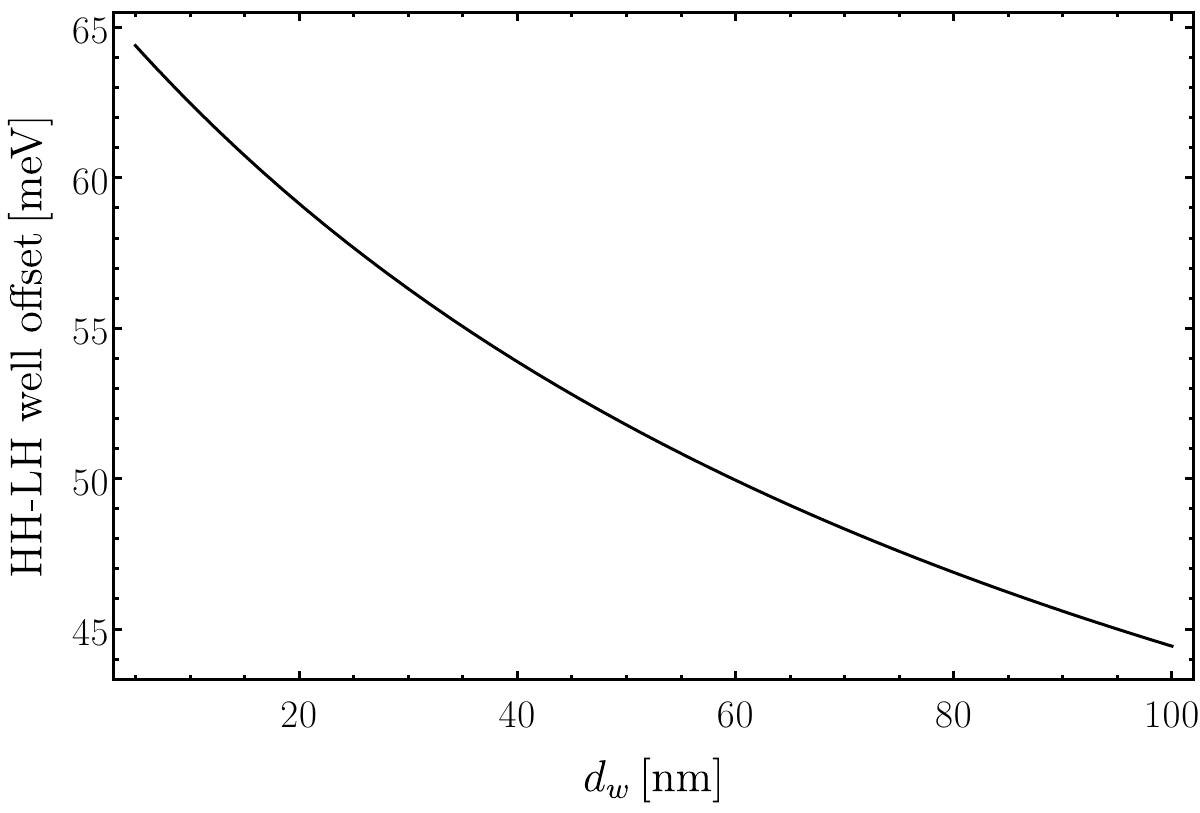}
    \label{subfig:balanced-offset}
    }
    \caption{\justifying \small
    A plot of the parabolic well (a) mixing, (b) $g_\perp$, and (c) HH-LH well minimum offset as a function of $d_w$ when strain balancing is included to the heterostructure. In contrast to Fig.~\ref{fig:mixing}, the correspondence between mixing and $g_\perp$ breaks down when the well is sufficiently wide, which occurs here at $d_w\approx20\,\text{nm}$. This is in part due to the decrease in HH-LH energy offset as a result of imposing Eq.~\eqref{eq:strain-balancing} as a constraint. 
    }
    \label{fig:strain-balance}
\end{figure*}
\end{widetext}

%%%%%%%%%%%%%%%%%%%%%%%%%%%%%%%%%%%%%%%%%%%%%%%%%%%

\section{Conclusion}
We have investigated the impact of quantum dot geometry and confinement profile on the effective g-factor of hole-spin qubits in strained Ge/SiGe heterostructures. Motivated by the intuition that isotropic wavefunctions yield isotropic g-factors, we carried out a detailed analysis using multiband envelope function theory. Simple quantum well models enabled us to predict the length scales required to suppress g-factor anisotropy. In particular, we find that a much stronger lateral confinement compared to state-of-the-art is necessary to produce a nearly isotropic geometry. Deviations from these predictions arise primarily from HH-LH mixing, which is sensitive to both the finite confinement potential and the strain profile. Although our predicted isotropic regime may not be feasible with the current state-of-the-art, we have shown that engineering confinement remains an effective strategy for reducing g-factor anisotropy.

The specific quantum well architecture plays a critical role in implementing this strategy. While both square and parabolic wells can, in principle, support isotropic wavefunctions with strong enough in-plane confinement, their handling of strain is qualitatively different. We have numerically shown that parabolic wells exhibit no asymptotic behavior as the well thickness is changed, allowing them to have out-of-plane scaling behavior that is not available to square wells. Furthermore, the wider profile allows the parabolic well to be grown with more relaxed strain requirements, making it less sensitive to strain-induced defects or imperfections. We find that our numerical g-factor calculations are highly sensitive to the $\eta_\text{LH}$, particularly in the quasi-2D dot limit, which suggests that the use of thin wells may not be ideal for addressing the yield uniformity problem. Our study of parabolic wells shows that engineering the g-factor properties through careful design of HH-LH mixing and strain dynamics may help in finding a viable alternative. Extending our analytical framework to incorporate interactions between in-plane and out-of-plane well basis states could help provide a better understanding of the competing physical effects when the in-plane and out-of-plane confinement lengths are comparable, especially in the strong confinement regime $(<10\,\text{nm})$.

Finally, we note that this study focused primarily on out-of-plane anisotropy. However, materials such as Ge exhibit strong anisotropy even within the plane of the heterostructure~\cite{Hendrickx_2024}. Our results are encouraging, as they suggest the possibility of also mitigating the in-plane anisotropy by carefully tuning the lateral confinement lengths $\sigma_{x}$ and $\sigma_{y}$. A complete analysis of in-plane anisotropy would require careful modeling of how HH-LH mixing, strain, finite confinement potentials, and noise influence the full 3D qubit geometry.

%%%%%%%%%%%%%%%%%%%%%%%%%%%%%%%%%%%%%%%%%%%%%%%%%%%

\begin{acknowledgments}

Research was sponsored by the Army Research Office and was accomplished under Cooperative Agreement Number W911NF-24-2-0082. The views and conclusions contained in this document are those of the authors and should not be interpreted as representing the official policies, either expressed or implied, of the Army Research Office or the U.S. Government. The U.S. Government is authorized to reproduce and distribute reprints for Government purposes notwithstanding any copyright notation herein.
This work was also supported by the Air Force Office of Scientific Research (AFOSR) through grant No. FA9550-23-1-0477. 
The computational results in this work were made possible by the Blackbird high performance computer cluster on NIST's Gaithersburg campus and supported by NIST's Research Services Office.  The authors acknowledge the Texas Advanced Computing Center (TACC) at The University of Texas at Austin for resources made available to NIST under contract number 1333ND25PNB180410 that have contributed to the research results reported within this paper. URL: https://tacc.utexas.edu

\end{acknowledgments}

%%%%%%%%%%%%%%%%%%%%%%%%%%%%%%%%%%%%%%%%%%%%%%%%%%%
%%%%%%%%%%%%%%%%%%%%%%%%%%%%%%%%%%%%%%%%%%%%%%%%%%%
%%%%%%%%%%%%%%%%%%%%%%%%%%%%%%%%%%%%%%%%%%%%%%%%%%%

\appendix
\renewcommand\thefigure{\thesection.\arabic{figure}} 
\renewcommand\thetable{\thesection.\arabic{table}} 

\section{Atomistic tight-binding calculations}
\label{app:tb}
\subsection{Computational methods} \label{subsec:tb-methods}

\paragraph*{}
We utilize an atomistic tight-binding (TB) model, similar to previous works \cite{linIncorporationRandomAlloy2019,linIncorporationRandomAlloy2024}, to aid in understanding the spatial structure of the wavefunction.  The barrier substrate above and below the well layer are comprised of a \ce{Si_{0.2}Ge_{0.8}} random alloy, generated with a pseudo-random number generator.  The well layer is entirely comprised of \ce{Ge} for the case of square wells.  For each parabolic well of a given width, a parabolic curve along $z$ is extrapolated between the edge of the 20\,\% \ce{Ge} barrier and 100\,\% \ce{Ge} at the center of the well.  The curve defines the target concentration of \ce{Ge} for a given atomic layer, and a random alloy is then generated for each $z$-value of the atomic lattice.

\paragraph*{}
The bare, zero-field TB Hamiltonian is
\begin{equation}
	H^0_{i,\alpha\alpha'}  = \delta_{\alpha\alpha'} \epsilon^0_{i,\alpha} + \Delta_{i,\alpha\alpha'} + \sum_{j=\text{nn}} t^0_{ij,\alpha\alpha'},
\end{equation}
where $i$ is the index for each individual atom (accounting for \ce{Si} or \ce{Ge}), $\alpha$ is index for the orbital-type ($spds^*$), and $j$ is summed over all nearest neighbors.  $\epsilon^0_{i,\alpha}$ is the on-site energy of orbital $\alpha$ at atom $i$, $\Delta_{i,\alpha\alpha'}$ represents the intrinsic spin-orbit coupling, and $t^0_{ij,\alpha\alpha'}$ is the hopping energy between orbital $\alpha$ of atom $i$ to orbital $\alpha'$ of atom $j$.

\paragraph*{}
The $sp3d5s^*$ orbital energies and couplings for \ce{Si} and \ce{Ge} are provided by Ref.~\cite{boykinValenceBandEffectivemass2004}; energies and couplings for \ce{Si_{0.5}Ge_{0.5}} are from Ref.~\cite{boykinBrillouinzoneUnfoldingPerfect2007}.  The alloyed lattice, including the well layer, is relaxed with a valence force field (VFF) method \cite{pryorComparisonTwoMethods1998}, and the TB Hamiltonian is adjusted accordingly to account for strain.  The VFF parameters $\alpha$, $\beta$, and the TB scaling exponents ($\eta$) for all three bond types are also found in Ref.~\cite{boykinBrillouinzoneUnfoldingPerfect2007}.

\paragraph*{}
The confining potential in the $z$-direction is given by the change in alloy concentration.  An additional constant electrical field ($F_z$) is applied in the $z$-direction to simulate the electrical bias of a plunger gate.  This is implemented in the TB Hamiltonian as a shift in the on-site orbital energies with respect to the electric potential at each atomic location.  The in-plane confinement field for the $x$ and $y$-directions are described with a parabolic electric potential, defined by a user-input characteristic length ($l_x$ and $l_y$).  The total on-site energy shift from electric fields is thus, in units of energy,
\begin{align}
	e\Phi_i(l_x, l_y, F_z) = \frac{\hbar^2}{2} \left( \frac{x_i^2}{m_\parallel l_x^4} + \frac{y_i^2}{m_\parallel l_y^4}  \right) + e F_z z_i,
\end{align}
where $x_i$, $y_i$, and $z_i$ are atomic positions for atom $i$.  For this paper, we choose $l_x=l_y$, and assume an $m_\parallel$ equal to the bulk \ce{Ge} light hole mass for the purposes of calculating the in-plane confining potential.

\subsection{Applied fields and effective orbital angular momentum} \label{subsec:tb-leff}

\paragraph*{}
Electric fields apply an on-site energy shift to the orbitals:
\begin{equation}
	\epsilon_{i,\alpha} \equiv \epsilon^0_{i,\alpha} - e\Phi_i,
\end{equation}
where as the magnetic field applies both an on-site  energy shift via the Zeeman term
\begin{equation}
	H^Z_{i,\alpha\alpha'}(\vec{B}) = \delta_{\alpha\alpha'} \mu_B (2\vec{S}_{i,\alpha}+ \vec{L}_{i,\alpha}) \cdot \vec{B}_i
\end{equation}
and an inter-orbital phase via the Peierls substitution
\begin{equation}
 	t_{ij,\alpha\alpha'} \equiv e^{ -\frac{ie}{\hbar} \int_{\vec{R}_{i}}^{\vec{R}_{j}} \vec{A}(\vec{s}) \cdot d\vec{s} } \, t^0_{ij,\alpha\alpha'},
\end{equation}
with $\mu_B$ the Bohr magneton, $\vec{S}_{i,\alpha}$ the spin operator, $\vec{L}_{i,\alpha}$ the atomic orbital operator, $\vec{A}$ the vector potential, and $\vec{R}_{i} (\vec{R}_{j})$ the position of atom $i (j)$.

\paragraph*{}
Perturbatively, our TB Hamiltonian can be written as
\begin{align}
	H_{i,\alpha\alpha'}(\vec{B}) =&\, \delta_{\alpha\alpha'} \epsilon_{i,\alpha} + \Delta_{i,\alpha\alpha'} + \sum_{j=\text{nn}} t_{ij,\alpha\alpha'} \nonumber \\
		&+ \delta_{\alpha\alpha'} \mu_B (2\vec{S}_{i,\alpha}+ \vec{L}_{i,\alpha}) \cdot \vec{B} \nonumber\\
	=&\, \delta_{\alpha\alpha'} \epsilon^0_{i,\alpha} -e \Phi_i + \Delta_{i,\alpha\alpha'} \nonumber \\
		&+ \sum_{j=\text{nn}} \left[ 1 - \frac{ie}{\hbar} \int_{\vec{R}_{i}}^{\vec{R}_{j}} \vec{A}(\vec{s}) \cdot d\vec{s} \right] \, t^0_{ij,\alpha\alpha'} \nonumber\\
		&+  \delta_{\alpha\alpha'} \mu_B (2\vec{S}_{i,\alpha}+ \vec{L}_{i,\alpha}) \cdot \vec{B} \nonumber\\
		&+ \mathcal{O}(\vec{B}^2)  \nonumber\\
	\approx&\, H^0_{i,\alpha\alpha'} -e \Phi_i + H^{\mathcal{O}(1)}_{i,\alpha\alpha'}(\vec{B}) + H^Z_{i,\alpha\alpha'}(\vec{B}), \nonumber\\
\end{align}
where
\begin{equation}
	H^{\mathcal{O}(1)}_{i,\alpha\alpha'}(\vec{B}) \equiv - \sum_{j=\text{nn}} \left[ \frac{ie}{\hbar} \int_{\vec{R}_{i}}^{\vec{R}_{j}} \vec{A}(\vec{s}) \cdot d\vec{s} \right] \, t^0_{ij,\alpha\alpha'}
\end{equation}
is the first order Peierls contribution.  The following results are obtained by partially diagonalizing $H\large|_{\vec{B}=0} = H^0 -e\Phi$ to obtain the spin-degenerate TB wavefunctions, which are then used as the perturbation basis for $H'\large|_{\mathcal{O}(\vec{B})} = H^{\mathcal{O}(1)}(\vec{B}) + H^Z(\vec{B})$ to obtain the magnetic splitting to first order.

\subsection{Tight-binding hole states g-factors} \label{subsec:tb-g_fac}

\paragraph*{}
For a constant magnetic field, we can further simplify the first order Peierls contribution as the effective spatial orbital angular momentum of the wavefunction dotted with the magnetic field, such that the magnetic splitting is a linear combination of three contributions:
\begin{align}
	H(\vec{B}) = H\large|_{\vec{B}=0} + \mu_B (\hat{\vec{L}}^\text{eff} + 2\hat{\vec{S}}+ \hat{\vec{L}}) \cdot \vec{B} + \mathcal{O}(\vec{B}^2),
\end{align}
where $\hat{\vec{S}}$ and $\hat{\vec{L}}$ represent the total spin and atomic orbital momentum, respectively.   The overall spatial angular momentum of the wavefunction ($\hat{\vec{L}}^\text{eff}$), as derived from the first order Peierls substitution, is strongly coupled to the magnetic spin response due to the large SOI in Ge dots.  This results in hole states that are strongly sensitive to the geometry of the dot and the direction of the magnetic field.  With strain, $\hat{\vec{L}}^\text{eff} \cdot \vec{B}$ is often the largest contribution to the total splitting.

\begin{figure}[ht!]
	\subfloat[22nm square well]{
		\includegraphics[width=0.24\textwidth]{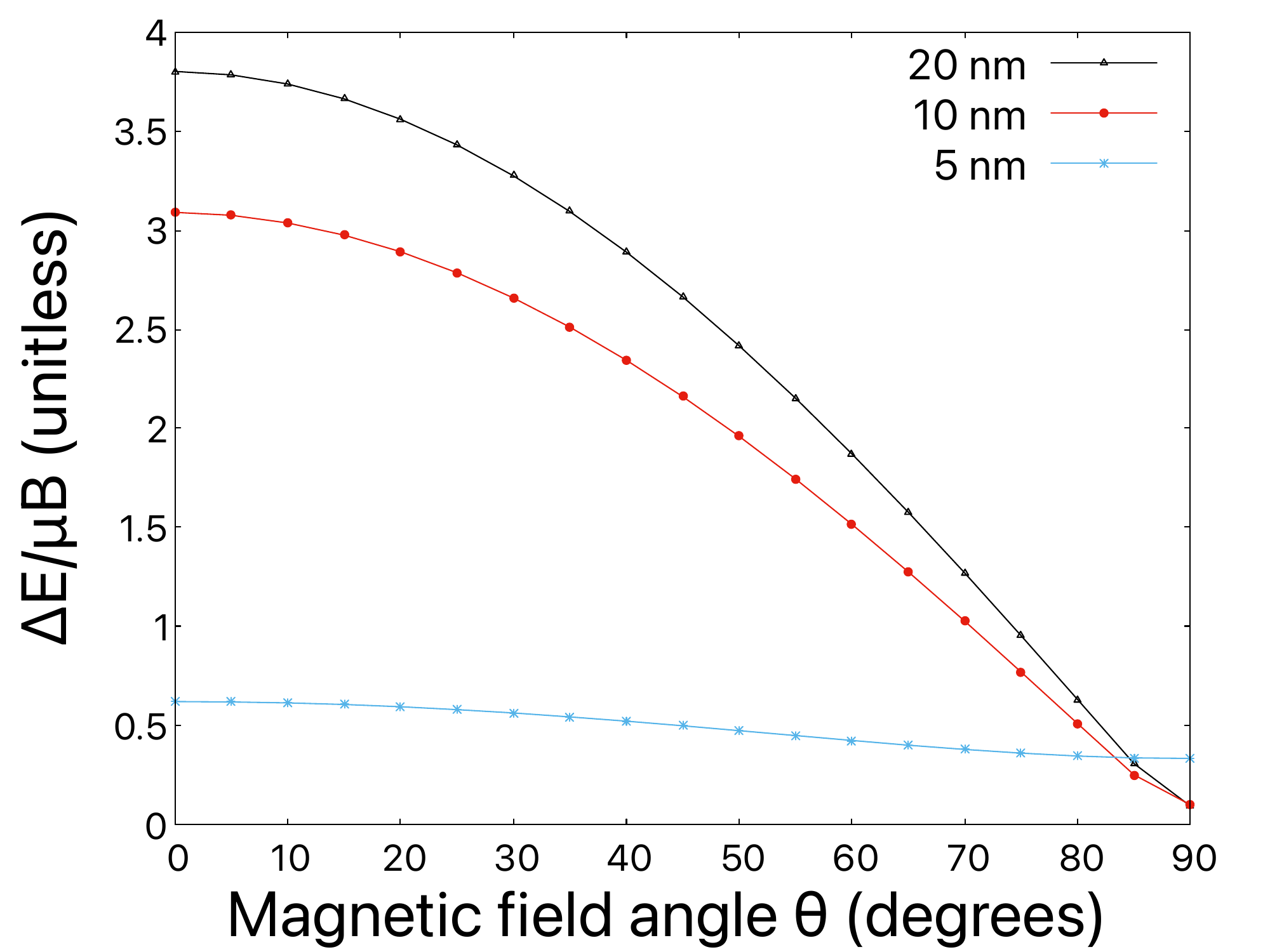}
		\label{fig-sub:tb-g_fac-sq}
	}
	\subfloat[22nm parabolic well]{
		\includegraphics[width=0.24\textwidth]{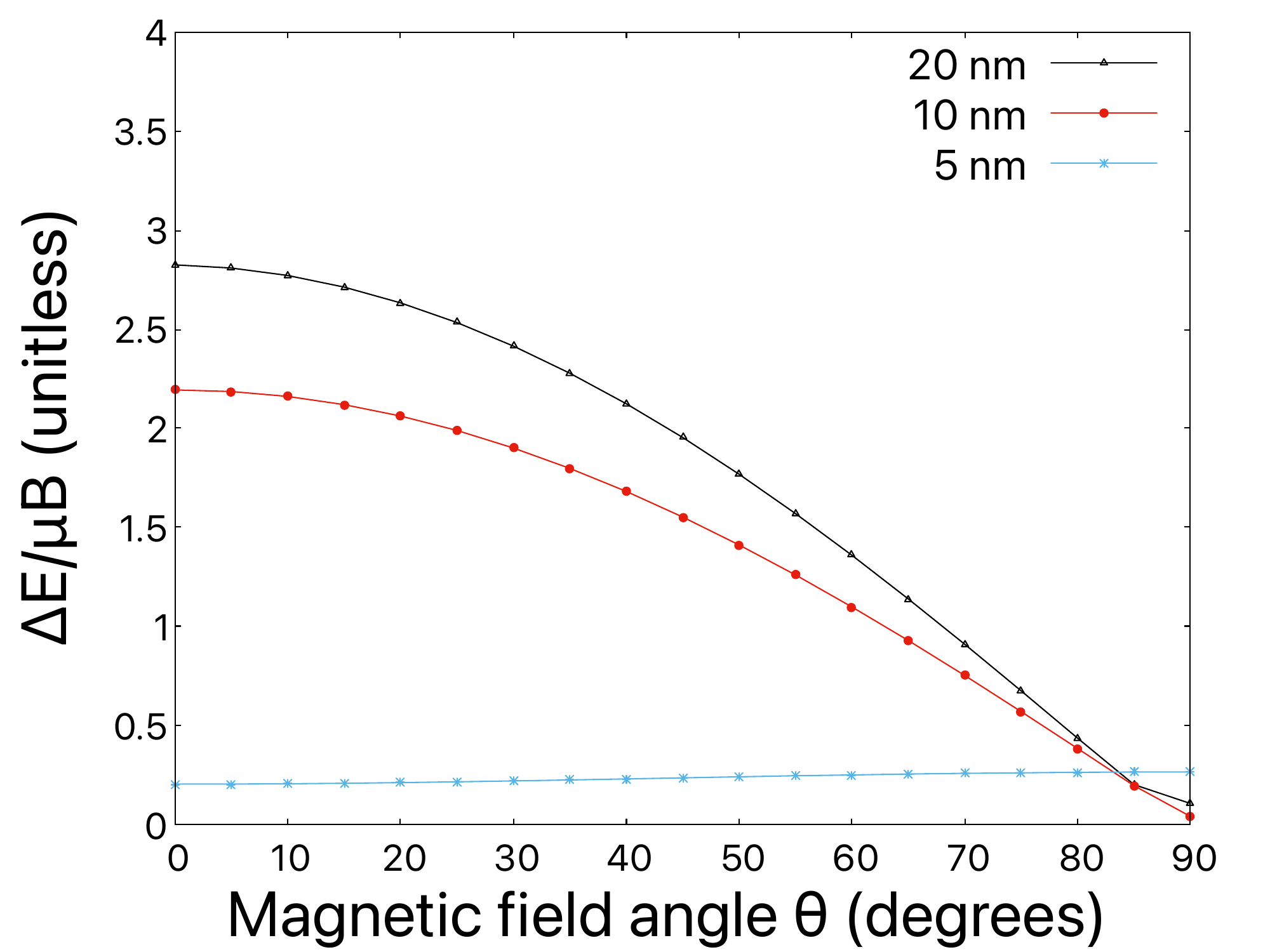}
		\label{fig-sub:tb-g_fac-para}
	}
	\caption{\justifying \small Magnetic splitting (g-factor) of the lowest energy hole state in a \SI{22}{\nano\meter} thick (a) square \ce{Ge} well (b) parabolic \ce{SiGe} well.  Plot labels indicate the input $l_{x/y}$, representing the strength of the in-plane electrostatic confinement.  $F_z = \SI{0.5}{\mega\volt\per\meter}$.  For both cases, we see strong anisotropy between the out-of-plane ($\theta=0^\circ$) and in-plane(($\theta=90^\circ$) splitting, with anisotropy greatly reduced at $l_{x/y}=\SI{5}{\nano\meter}$ due to a comparable in-plane and out-of-plane wavefunction width (Fig~\ref{fig:tb-rho}).}
	\label{fig:tb-g_fac}
\end{figure}

\paragraph*{}
In Figure~\ref{fig:tb-g_fac}, we plot the magnetic splitting \SI{22}{\nano\meter} thick wells as a function of the magnetic field angle for various strengths of lateral confinement.  For the more realistic lateral confinement strengths, $l_{x/y}$, of \SI{20}{\nano\meter} or \SI{10}{\nano\meter}, both the square and the parabolic wells show a strong (weak) magnetic response when the magnetic field is out-of-plane (in-plane), with the stronger lateral confinement showing a lower anisotropy between the out-of-plane and in-plane responses.  For these confinement strengths, the wavefunctions are wider in-plane than they are
tall out-of-plane, resulting in an $\vec{L}^\text{eff} $ that points strongly in the out-of-plane direction, regardless of magnetic field direction.  The spin also picks up this directional preference of the spatial angular momentum via the SOI, further compounding the strong anisotropy.

\begin{figure}[ht!]
	\subfloat[22nm square well]{
		\includegraphics[width=0.24\textwidth]{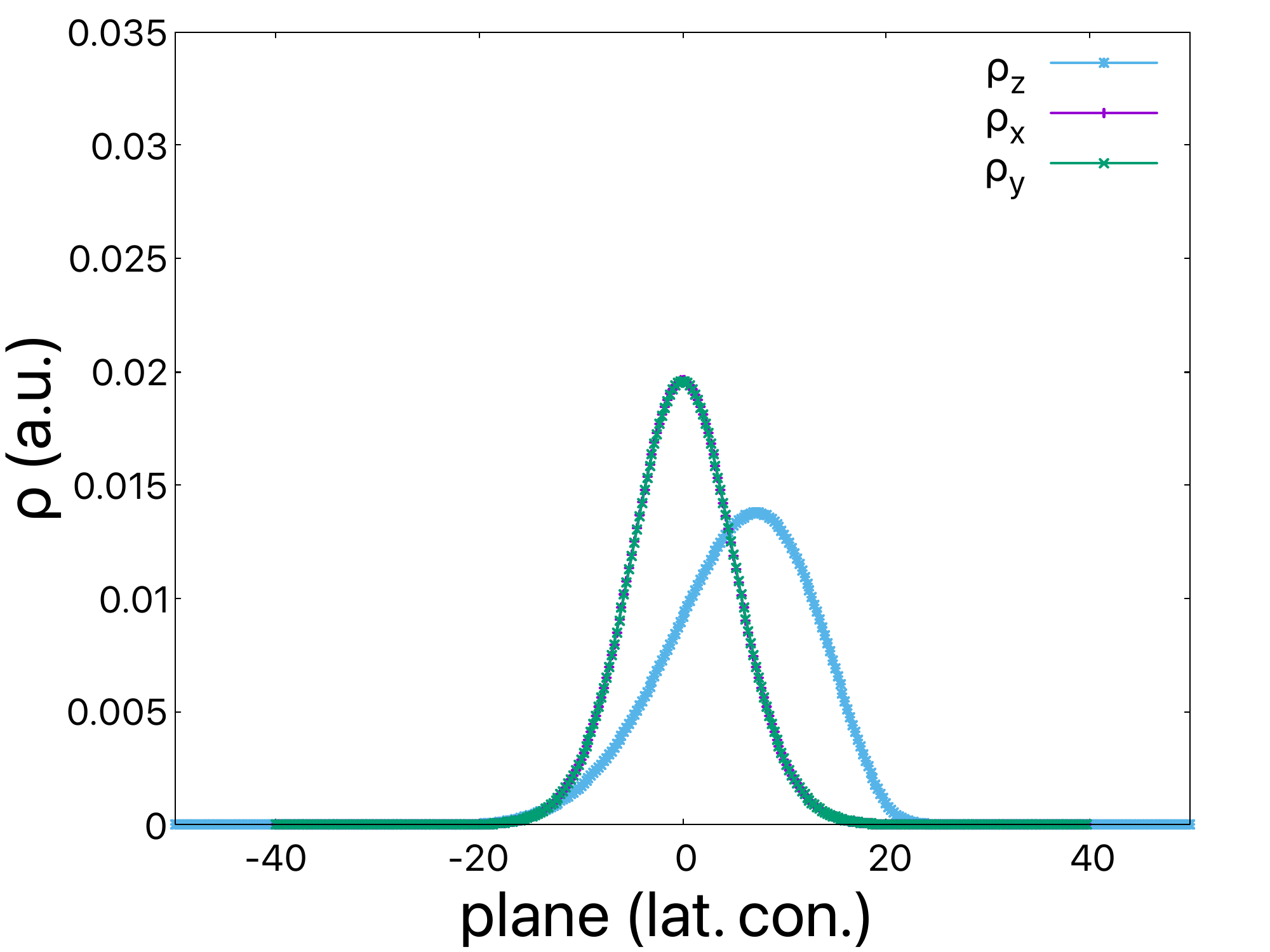}
		\label{fig-sub:tb-rho-sq}
	}
	\subfloat[22nm parabolic well]{
		\includegraphics[width=0.24\textwidth]{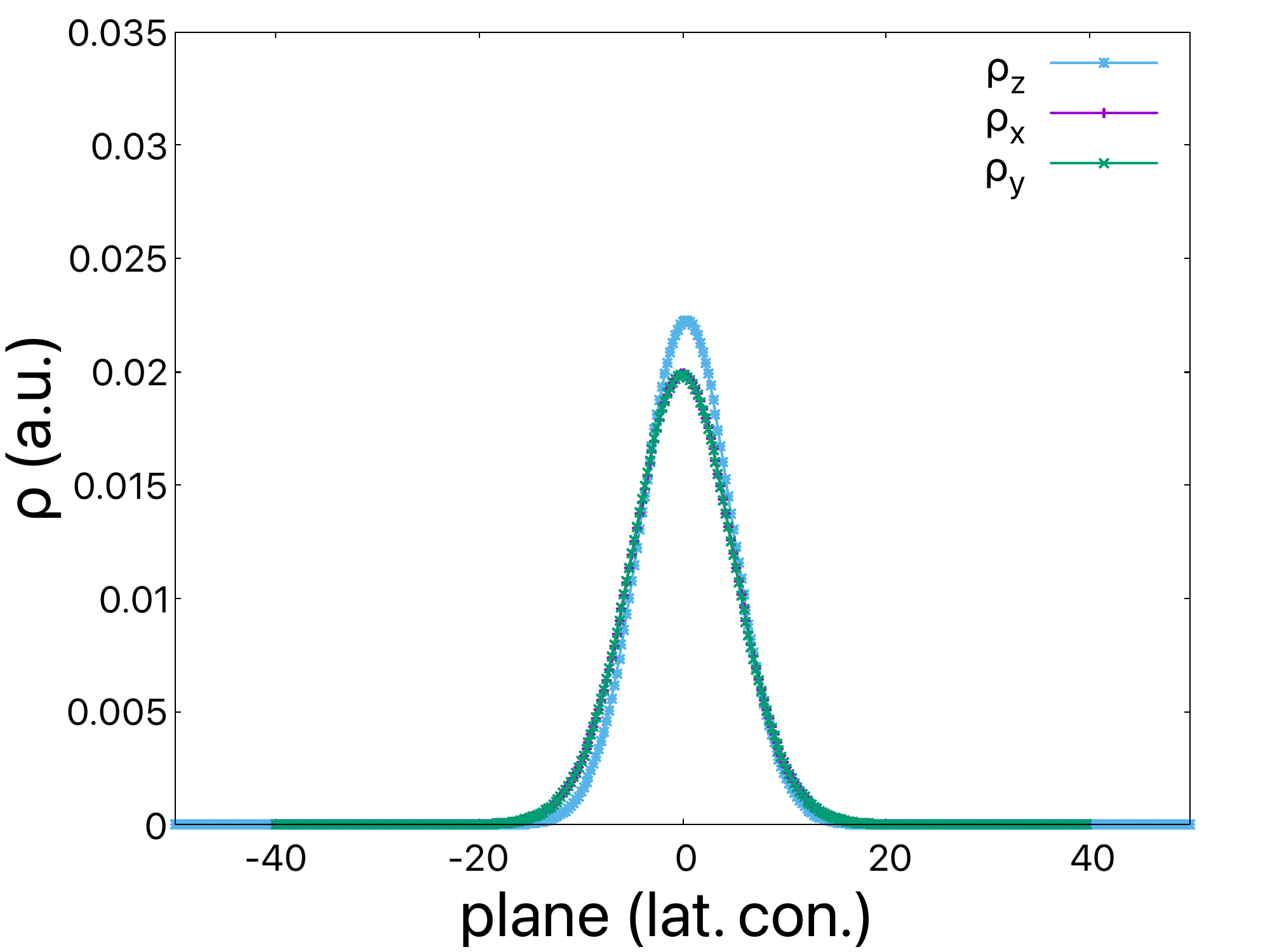}
		\label{fig-sub:tb-rho-para}
	}
	\caption{\justifying \small Charge distribution of (a) square \ce{Ge} well (b) parabolic \ce{SiGe} well, with $l_{x/y}=\SI{5}{\nano\meter}$.  $F_z = \SI{0.5}{\mega\volt\per\meter}$.}
	\label{fig:tb-rho}
\end{figure}

\paragraph*{}
As we increase the lateral confinement to a very strong $l_{x/y} = \SI{5}{\nano\meter}$, the width of the wavefunction becomes comparable for the in-plane and out-of-plane directions (Fig~\ref{fig:tb-rho}).  The result is a decrease in the $\hat{\vec{L}}^\text{eff} \cdot \vec{B}$ contribution to magnetic splitting, and, more importantly, the spatial orbital momentum of the wavefunction is no longer locked to the out-of-plane direction due to the flat disk-shaped geometry of the dot. We thus observe a removal of anisotropy as both the spin and spatial angular momentum are allowed to freely rotate with the direction of the magnetic field.

\subsection{Drop in g-factor for thin wells} \label{subsec:tb-11nm}

\begin{figure}[ht!]
	\subfloat[11nm square well]{
		\includegraphics[width=0.24\textwidth]{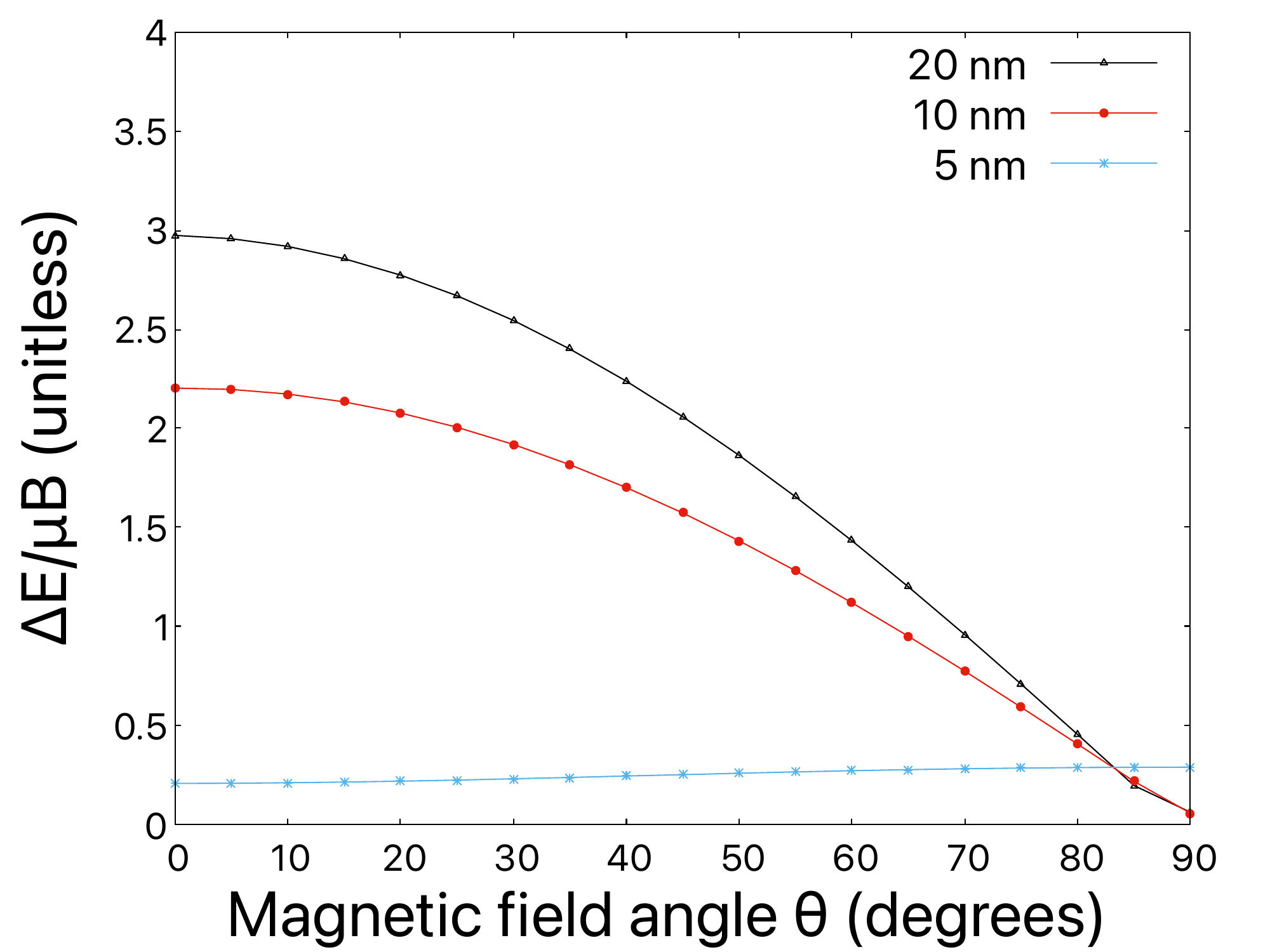}
		\label{fig-sub:tb-10nm-sq}
	}
	\subfloat[11nm parabolic well]{
		\includegraphics[width=0.24\textwidth]{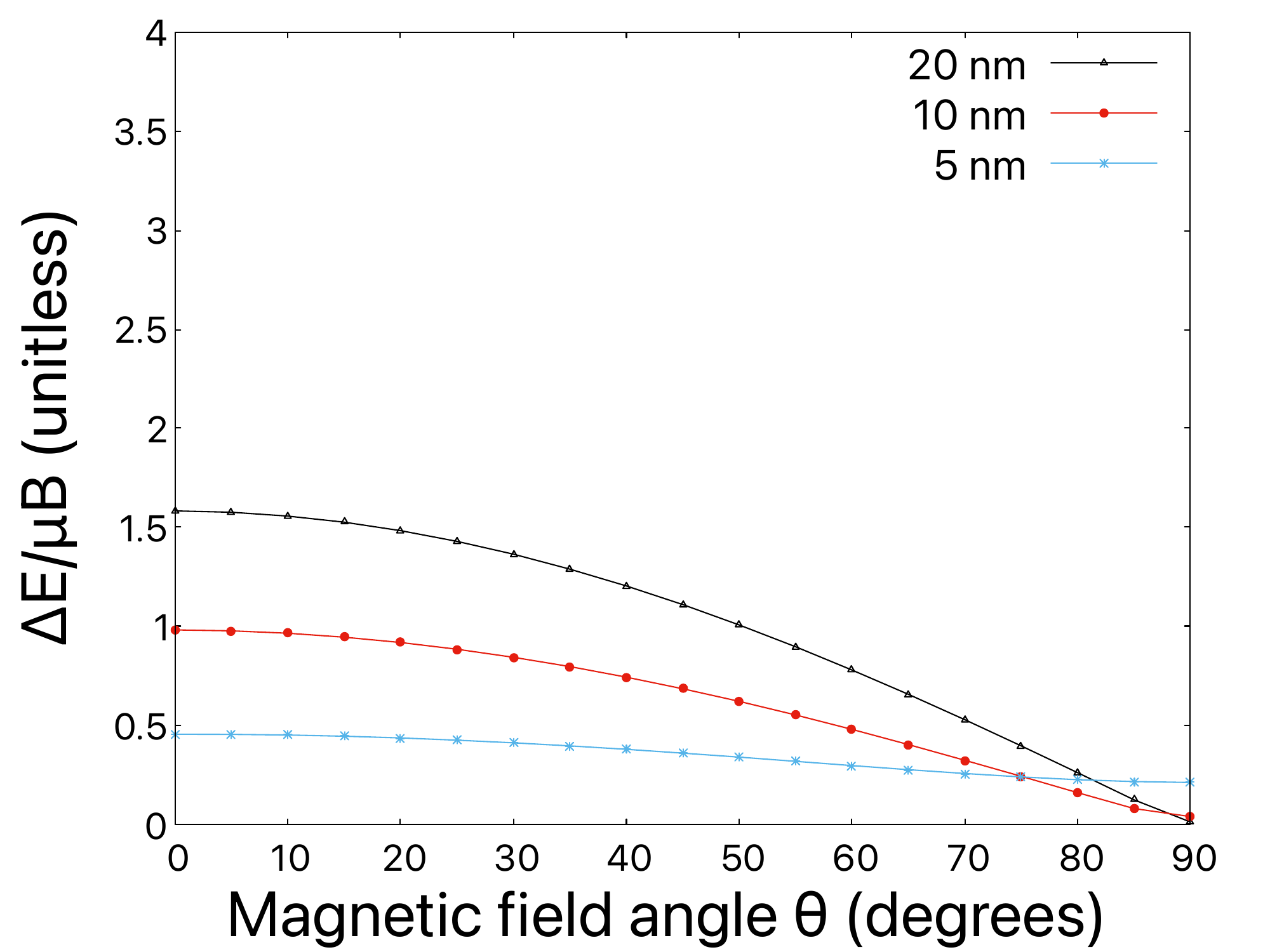}
		\label{fig-sub:tb-10nm-para}
	}
	\caption{\justifying \small Magnetic splitting (g-factor) of the lowest energy hole state in a \SI{11}{\nano\meter} thick (a) square \ce{Ge} well (b) parabolic \ce{SiGe} well.  Plot labels indicate the input $l_{x/y}$, representing the strength of the in-plane electrostatic confinement.  $F_z = \SI{0.5}{\mega\volt\per\meter}$.  For very thin wells, g-factor is reduced due to a lowered contribution from the Peierls term.}
	\label{fig:tb-10nm}
\end{figure}

\paragraph*{}
Another method of reducing the effect $\vec{L}^\text{eff}$ has on the magnetic field is the breakup of the spatial orbital structure at very thin wells.  As Fig~\ref{fig:tb-10nm} shows, a thin \SI{11}{\nano\meter} well results in a overall reduction in g-factor as the $\hat{\vec{L}}^\text{eff} \cdot\vec{B}$ contribution is lowered.  Since the out-of-plane splitting is disproportionately lowered, we see an overall reduction in anisotropy.  However, the spatial angular momentum is now very sensitive to strain, unnecessarily complicating the picture if the same reduction of g-factor anisotropy can be achieved at larger wells.

\paragraph*{}
In conclusion, the strong g-factor anisotropy of hole states in \ce{Ge} wells is a result of a strong out-of-plane spatial angular momentum coupling to the magnetic spin response with a strong SOI. Without modifying the SOI strength, the strong out-of-plane spatial angular momentum can be reduced by either going to wider wells (such that it no longer has a preference for the out-of-plane direction) or by very thin wells (such that the spatial orbital structure of the hole states is broken up).  However, the latter is a much less reliable method, as the g-factor of thin wells is also highly dependent on a myriad of other factors.  Thus, we recommend widening the wells to approach bulk \ce{Ge}, with parabolically shaped wells being an experimentally viable method of tackling such a problem. 

\bibliography{bib_2025-03-05}
\bibliographystyle{apsrev4-2}

\end{document}